\documentclass{aa}
\usepackage{graphicx}
\usepackage{amsmath}
\usepackage{amssymb}
\usepackage{color}
\usepackage{natbib}
\usepackage{hyperref}
\bibpunct{(}{)}{;}{a}{}{,}

\begin{document}

\title{Three-lobed near-infrared Stokes $V$ profiles in the quiet Sun}

\author{Christoph Kiess
  \and Juan Manuel Borrero \and Wolfgang Schmidt}
\authorrunning{Kiess, Borrero, Schmidt}
\titlerunning{Three-lobed Stokes $V$ Profiles}
\institute{Kiepenheuer-Institut f\"ur Sonnenphysik, Sch\"oneckstr. 6, D-7104 Freiburg, Germany\\
\email{[kiess, borrero, schmidt]@leibniz-kis.de}}

\abstract
{The 1.5-m GREGOR solar telescope can resolve structures as small as 0.4" at near-infrared (NIR) wavelengths on the Sun. 
At this spatial resolution the polarized solar spectrum shows complex patterns, such as large horizontal 
and/or vertical variations of the physical parameters in the solar photosphere.
}
{We investigate a region of the quiet solar photosphere exhibiting three-lobed Stokes $V$ profiles
in the Fe \textsc{I} spectral line at 15648 {\AA}. The data were acquired with the 
GRIS spectropolarimeter attached to the GREGOR telescope. We aim at investigating the thermal, 
kinematic and magnetic properties of the atmosphere responsible for these measured complex signals.
}
{The SIR inversion code is employed to retrieve the physical parameters of the lower solar photosphere from the observed polarization signals.
We follow two different approaches. On the one hand, we consider that the multi-lobe circular polarization signals
are only produced by the line-of-sight variation of the physical parameters. We therefore invert the data assuming a single atmospheric
component that occupies the entire resolution element in the horizontal plane and where the physical parameters vary with
optical depth $\tau$ (i.e., line-of-sight). On the other hand, we consider that the multi-lobe circular polarization signals are produced
not by the optical depth variations of the physical parameters but instead by their horizontal variations. Here we invert the data
assuming that the resolution element is occupied by two different atmospheric components where the kinematic
and magnetic properties are constant along the line-of-sight.
}
{
Both approaches reveal some common features about the topology responsible for the 
observed three-lobed Stokes $V$ signals: both a strong ($>1000$ Gauss) and a very weak ($< 10$ Gauss) magnetic field
with opposite polarities and harboring flows directed in opposite directions must co-exist 
(either vertically or horizontally interlaced) within the resolution element.
}
{} 

\maketitle


\section{Introduction}

Highly asymmetric circular polarization signals emerging from 
the solar surface and deviating from their
usual anti-symmetric shape are known to be caused by variations along the line-of-sight in the magnetic and kinematic
properties of the solar plasma \citep{auer78ncp,landi81ncp,landi83ncp,landolfi1996ncp}, as well as by the presence of structures that remain 
horizontally unresolved at the spatial resolution of the observations. Two examples of such signals 
are single- and three-lobed Stokes $V$ profiles. While the former can only be explained by gradients along the line-of-sight in
the kinematic and magnetic parameters, the latter can also be produced by the presence of structures that
remain unresolved at the spatial resolution of the observations. As the spatial resolution of the observations 
increases, unresolved structure becomes an increasingly unlikely explanation for three-lobed circular polarization signals.
In sunspots, these kinds of circular polarization profiles are relatively common at low resolution (i.e., $\approx 1"$), 
both in the visible \citep{Sanchez_Almeida_1992,solanki1993ncp, borrero2006} and near-infrared (NIR) \citep{ruedi1998,Toro_Iniesta_2001,Borrero_2004,Bellot_Rubio_2004,Borrero_2005} spectral lines. 
Moreover, many comparative and statistical studies between both spectral bands have been carried out at this angular resolution 
\citep{rolf2002ncp1,rolf2002ncp2,mueller2002ncp,borrero2007ncp,borrero2010ncp}. At high-spatial resolution (i.e., $< 0.5"$) similar 
studies have been conducted in visible \citep{Ichimoto_2007,ichimoto2008ncp,Franz_2013,Pozuelo_2016}, 
and NIR spectral lines \citep{Franz_2016}.\\

In the quiet Sun (i.e., internetwork, network, plage, faculae, etc.), most of the studies carried out at
low- and mid spatial resolution have shown only slightly asymmetric circular polarization profiles 
\citep{stenflo1984da,jorge1988ncp,jorge1989ncp,bellot2000ncp,Khomenko_2003}. A number of works, however, also 
found highly asymmetric Stokes $V$ profiles at this resolution in visible \citep{Socas_Navarro_2005}
and NIR \citep{Bellot_Rubio_2001} spectral lines, and concluded that they were the signature of shock fronts. 
As the spatial resolution of the observations has improved, so has the amount of single- and three-lobed Stokes $V$ profiles 
observed in the quiet Sun. In particular, a myriad of studies on these kinds of profiles have been carried out in the past few years
using spectral lines in the visible range. \citet{Viticchie_2012} and \citet{Quintero_Noda_2014a} identify single-lobed Stokes $V$ 
profiles as the signature of magnetic $\Omega$-loop structures, whereas \citet{Sainz_Dalda_2012} ascribe them to flux emergence and
submergence processes. Strongly blueshifted single-lobed Stokes $V$ signals were analyzed by \citet{Borrero_2010},
\citet{Martinez_Pillet_2011}, \citet{Quintero_Noda_2013}, and \citet{Borrero_2013} who associated them with supersonic upflows, 
possibly related to magnetic reconnection. \citet{Jafarzadeh_2015} found that these supersonic upflows could also manifest themselves in
three-lobed Stokes $V$ profiles. Highly asymmetric circular polarization profiles, related to downflows, have been studied by 
\citet{Quintero_Noda_2014b}.\\

All of the aforementioned studies of highly asymmetric Stokes $V$ profiles at high spatial resolution employed spectral lines 
in the visible range. Similar investigations in the NIR wavelength range, in particular around 15000 {\AA}, have the potential
of greatly contributing to our understanding of the interplay between convective motions, radiation, and magnetic fields in the
quiet Sun. This is a consequence of their greater sensitivity to magnetic fields compared to their visible counterparts. In addition,
spectral lines close to the 15000 {\AA} wavelength range sample layers of the solar photosphere that are about 60-80 km deeper
than the layers sampled by visible spectral lines \citep{chandra1946hminus,Borrero_2016}, thereby providing additional information
about the physical processes taking place in the quiet Sun.\\

Nevertheless, high-resolution observations in the NIR require telescopes with larger apertures. Fortunately, with the
infrared spectropolarimeter GRegor Infrared Spectrograph GRIS \citep{Collados_2012}, 
attached to the 1.5-m GREGOR solar telescope \citep{Schmidt_2012},
it is now possible to obtain spectropolarimetric observations (i.e., Stokes vector) of the quiet Sun with an angular resolution 
similar to the aforementioned studies with visible spectral lines (i.e., $\approx 0.4"$) but with spectral lines located at 15650 {\AA} 
\citep{marian2016gregor,Lagg_2016}. In this work we report on the detection, analysis, and interpretation of highly asymmetric (i.e., three-lobed) 
Stokes $V$ profiles in the Fe {\sc I} 15648 {\AA} spectral line in the quiet Sun. Sections~\ref{section:observations}
and ~\ref{section:verify} present such observations and data calibration, respectively. The data is then subjected to
a first analysis in Section~\ref{section:analysis} and a more detailed analysis (based on the inversion of the radiative transfer equation)
in Section~\ref{section:inversion}, using two different geometrical models. The results for each of the two models 
are presented in Section ~\ref{section:results} and discussed in Section~\ref{section:discussion}. Finally, Section~\ref{section:conclusions}
presents our conclusions and ideas for further studies.\\

\begin{table*}
\begin{center}
\caption{Atomic parameters of the used spectral lines.  \label{tab:lines}}
\begin{tabular}{cccccccc}
\hline
${\lambda_0}$\tablefootmark{b} & Specie & Electron Conf.\tablefootmark{b} & $g_\textrm{eff}$ & 
${\chi_\textrm{low}}$\tablefootmark{b} & $\log(gf)$ & $\sigma$ & $\alpha$ \\
\textrm{[\AA]} & & & & \textrm{[eV]} & & [$a_0^2$] & \\
\hline
15648.518 &  Fe\,I & ${^{7}D_1} - {^{7}D_1}$               &  3.0 & 5.426 & -0.669\tablefootmark{b} & 975\tablefootmark{b} &  0.229\tablefootmark{b} \\
15652.874 &  Fe\,I & ${^{7}D_5} - {^{7}D_{4.5}} {f[3.5]^0}$ & 1.45 & 6.246 & -0.095\tablefootmark{b} & 1427\tablefootmark{b} & 0.330\tablefootmark{b} \\
15662.018 &  Fe\,I & ${^{5}F_5} - {^{5}F_4}$               & 1.5  & 5.830 & 0.190\tablefootmark{c} & 1197 \tablefootmark{c}& 0.240\tablefootmark{c}\\
\hline
\end{tabular}
\tablefoot{$\sigma$ and $\alpha$ represent the cross-section (in units of Bohr's radius squared $a_0^2$) and velocity parameter of the atom undergoing 
the transition, respectively, for collisions with neutral atoms under the ABO theory \citep{abo1,abo2,abo3}\tablefootmark{a}{Values taken from \citet{Nave_1994}}. 
\tablefootmark{b}{Values taken from \citet{Borrero_2003}}.\tablefootmark{c}{Values taken from \citet{Bloomfield_2007}}}
\end{center}
\end{table*}

\section{Observations and data calibration}
\label{section:observations}

The data employed in this work were acquired with the 1.5-m GREGOR solar telescope \citep{Schmidt_2012}, 
located at El Teide Observatory on Tenerife, Spain. The GRegor Infrared Spectrograph \citep[GRIS ;][]{Collados_2012} 
was employed to record the Stokes vector $(I,Q,U,V)$ of several
Fe \textsc{I} spectral lines located at 15650 {\AA}. The atomic parameters of the spectral lines
selected for analysis within the aforementioned spectral range are given in Table ~\ref{tab:lines}.
The sampling along the wavelength axis is 40 m{\AA} per pixel. The spatial direction along the slit
samples 61.4" in 450 pixels. The scan was performed perpendicularly to the slit, with a step size of 0.135" and 
a total of 99 steps, resulting in a field-of-view with a size of 61.4$\times$13.4 arcsec$^2$.
The exposure time for each slit position was 4.8 seconds, resulting in a noise level in the polarization signal
($Q$, $U$, and $V$) of $4.5 \cdot 10^{-4}$ in units of the quiet Sun continuum intensity.\\

\begin{figure}[!h]
\resizebox{\hsize}{!}{\includegraphics{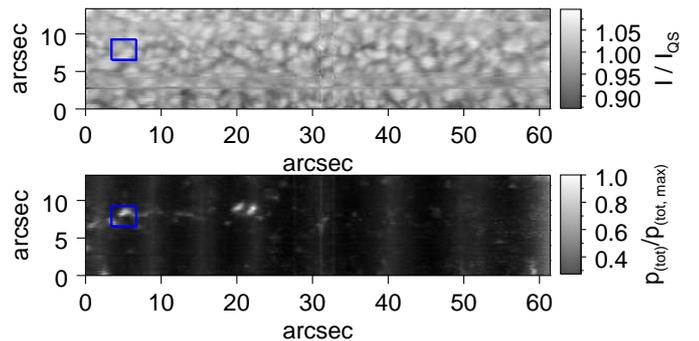}}
\caption{The upper panel shows the continuum intensity (normalized to the average value over the field-of-view)
of the scan used in this paper. The lower panel is the total polarization normalized to its maximum.
Scan direction is from bottom to top. The horizontal axis corresponds to the direction of the slit.
The blue rectangle marks the area analyzed in this paper, also seen in Fig.~\ref{fig:continuummap_contour} (top panel).}
\label{fig:continuum}
\end{figure}

Figure~\ref{fig:continuum} shows a map of the continuum intensity (top) and total polarization (bottom) 
reconstructed from the individual slit positions. Residual fringes not completely removed during the data
reduction are visible as periodic vertical stripes in the total polarization map \citep[see][for details]{Franz_2016}.
The region marked by the blue square, located at
approximately $(x,y) \approx (5",8")$, is the region selected for further analysis in this paper.
The horizontal stripes visible in Fig.~\ref{fig:continuum} are short jumps of the tip-tilt mirror of the 
adaptive optic system \citep{Berkefeld_2012}. Fortunately, during the scan of the region of interest, 
the adaptive optics system worked stably. In this region the spatial resolution is estimated, from the 
power-spectrum, to be about 0.4 arcsec.\\

Standard calibration routines were applied to the raw data from the GRIS instrument. The calibration steps include:
cropping, correction for dark current, flat-field, polarimetric calibration and wavelength calibration\footnote{The
wavelength calibration is done assuming that the line-core positions of the spectral lines on the 
average quiet Sun intensity (Stokes $I$) profiles are located at the laboratory wavelengths in Table~\ref{tab:lines}
and shifted by 395 ms$^{-1}$, which is the estimated net effect of the convective blueshift and gravitational redshift. 
This estimation is done with the FTS spectrum.}. These so-called level~1 data are, after a proprietary phase, publicly available at 
\href{http://archive.leibniz-kis.de/pub/gris/}{http://archive.leibniz-kis.de/pub/gris/}. Level~1 data can be used for 
quick-look purposes and some spectral analysis, but need further calibration before more reliable physical parameters can 
be derived from them via inversion. Some of these additional calibration steps have been described elsewhere:
spectral veil correction and determination of spectral point-spread-function (PSF), reduction of 
spectral fringes and wavelength calibration \citep[][]{Franz_2016,Borrero_2016}. Here we would like to focus our attention on the 
issue of the continuum normalization for the intensity profile (Stokes $I$).\\

\begin{figure}
\resizebox{\hsize}{!}{\includegraphics{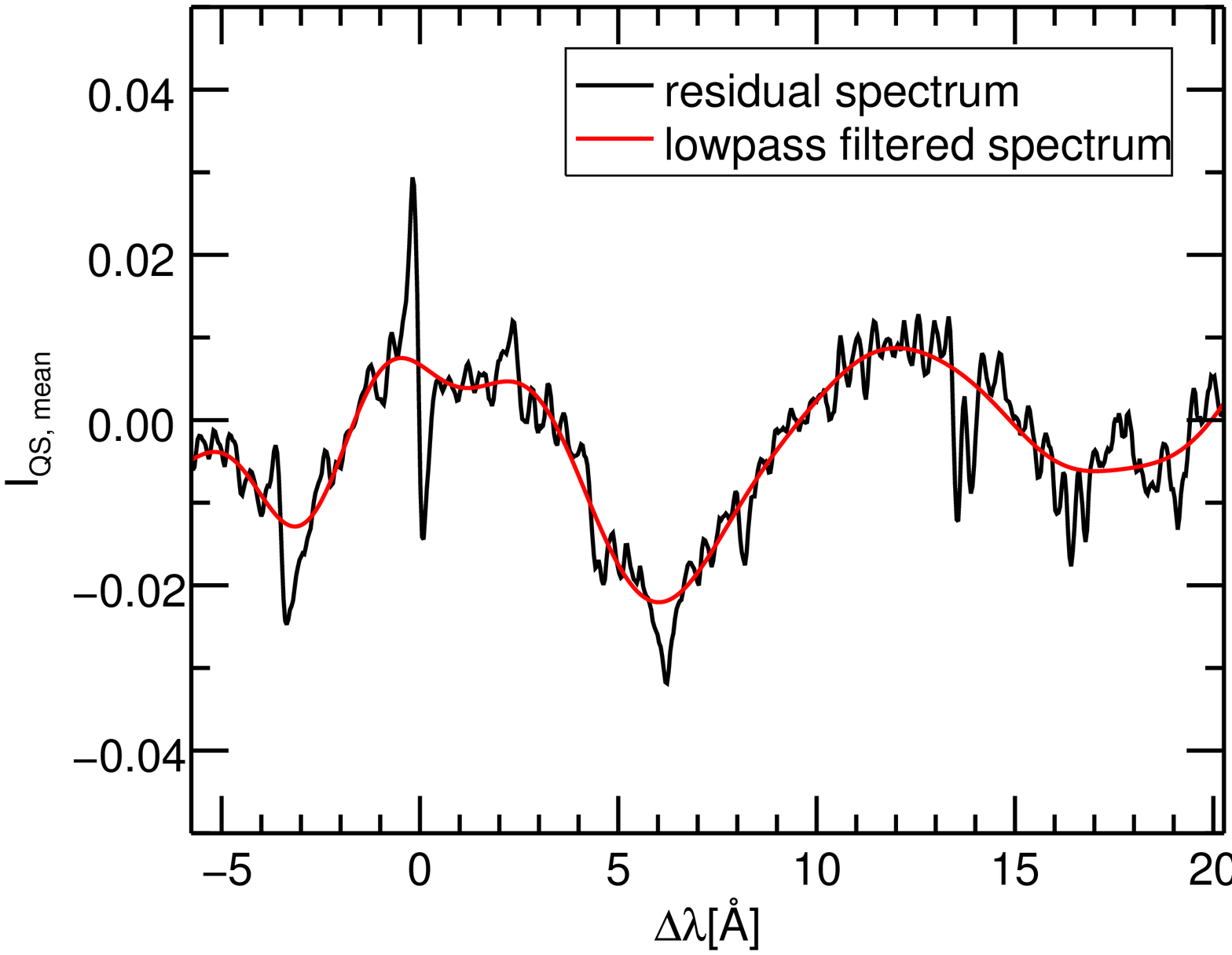}} 

\resizebox{\hsize}{!}{\includegraphics{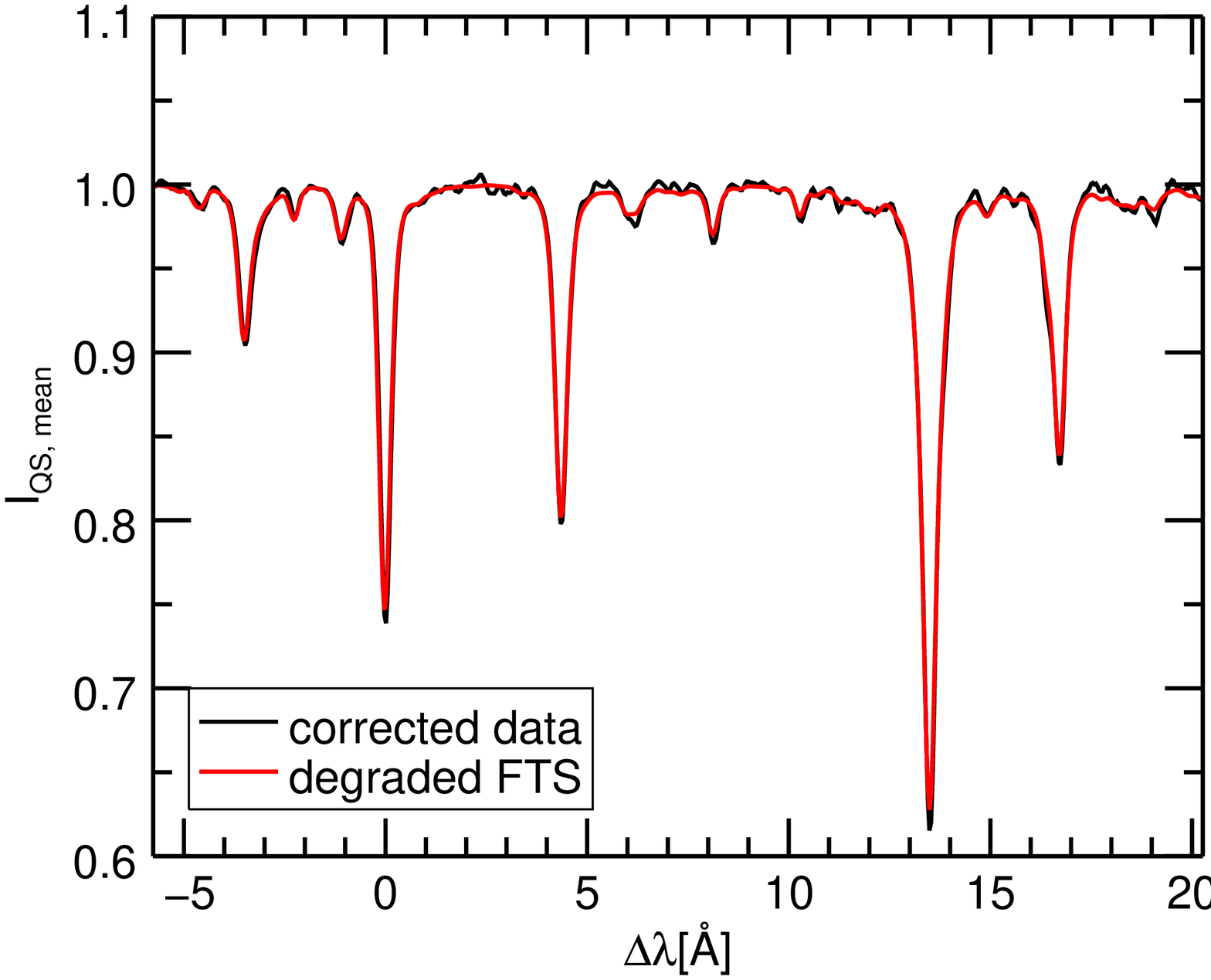}}
\caption{Continuum calibration using a comparison between the FTS spectrum and the measured spectrum. 
The top panel shows the residual spectrum (black) and the lowpass filtered curve used for the calibration (red).
In the bottom panel we compare the degraded FTS spectrum (red) with the calibrated spectrum (black). $\Delta \lambda = 0$
refers to the Fe {\sc I} 15648.5 {\AA}~ spectral line in Table~\ref{tab:lines}.}
\label{fig:continuum_fts_fft}
\end{figure}

The measured continuum intensity is not constant with wavelength but varies by a few per cent over the observed spectral 
range \citep[see Fig.~1 in][]{Franz_2016}. To detrend the continuum, we compared the average Stokes $I$ of the measured data
(i.e., quiet Sun) to that of the Fourier Transform Spectrograph \citep[FTS,][]{Livingston_Wallace_1991} at the same wavelength range. 
For this comparison to be meaningful, we must first degrade the FTS spectrum to the same spectral resolution of the GRIS instrument 
by applying the instrument's spectral PSF \citep[a Gaussian function with $\sigma = 70$ m{\AA} and 12 \% spectral veil, as in][]{Borrero_2016}
\footnote{The same PSF is also used during the inversions (Sect.~\ref{section:inversion}).}. We then subtract both spectra and fit the residual 
spectrum with a Fourier-based method. This procedure is showcased in Fig.~\ref{fig:continuum_fts_fft} (top panel), where the residual 
spectrum and Fourier-filtered curve are shown in black and red, respectively. To normalize the dataset, we divide it by the Fourier-filtered 
curve and renormalize the continuum to one. In order to highlight the reliability of this approach, we compare, in the bottom panel of 
Fig.~\ref{fig:continuum_fts_fft}, the degraded FTS spectrum (red) with the average Stokes $I$ over the recorded field-of-view after 
the continuum normalization is performed (black). As can be seen, both spectra have a very similar continuum.\\

\section{Verifying data calibration}
\label{section:verify}

\begin{figure}
\resizebox{\hsize}{!}{\includegraphics{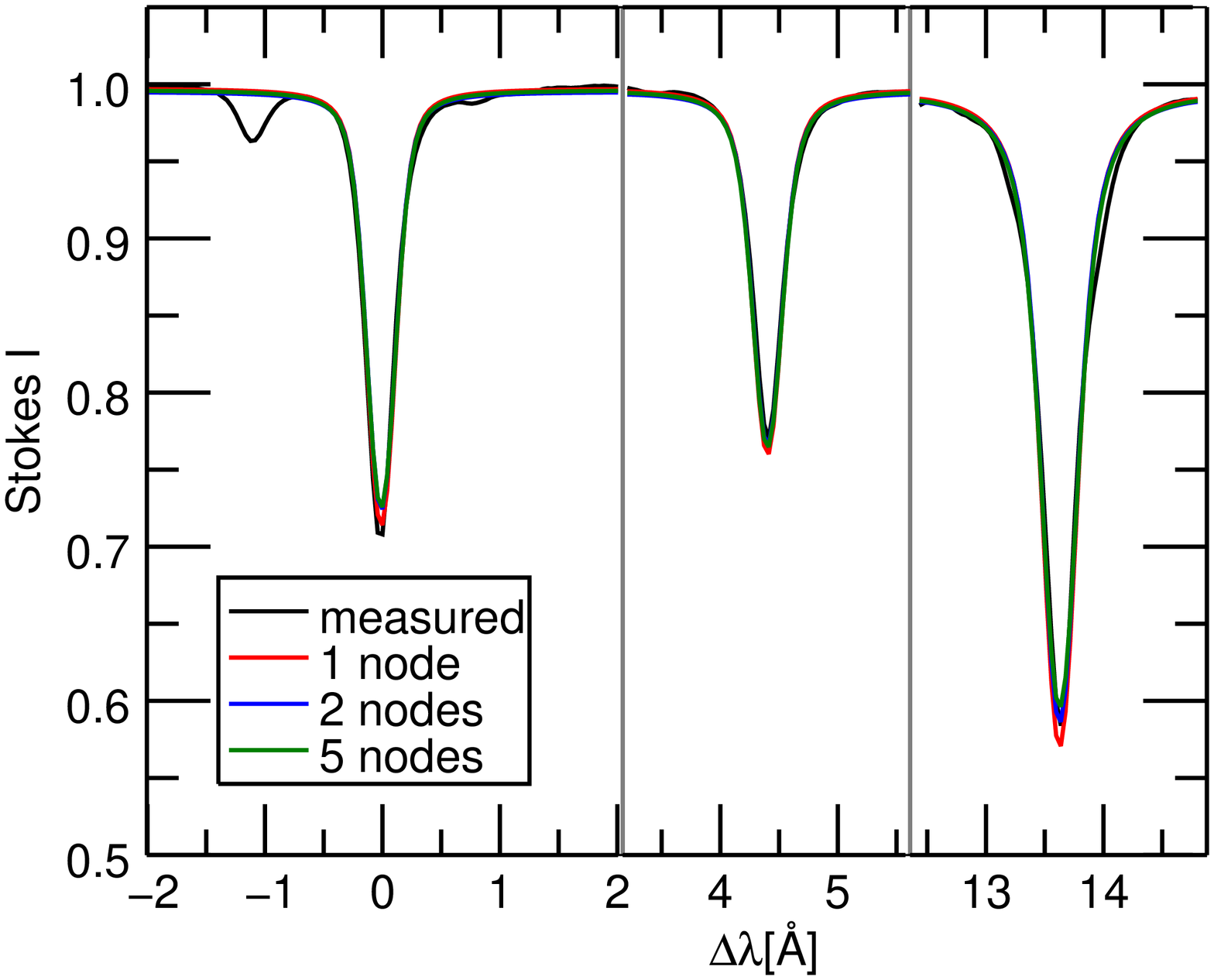}} 

\resizebox{\hsize}{!}{\includegraphics{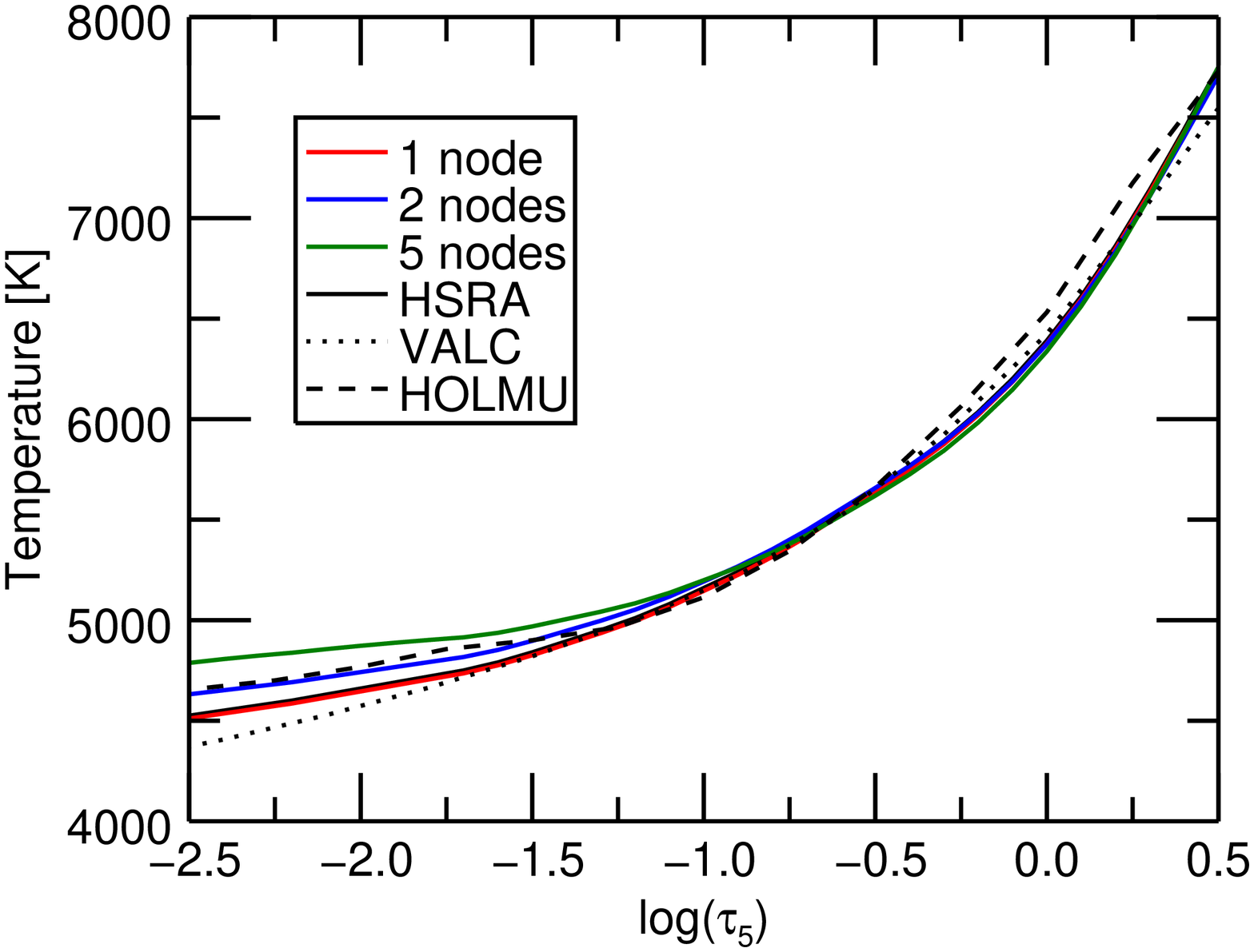}}
\caption{Calibration verification. All curves refer to the mean quiet Sun spectrum. {\it Top panel}: 
observed Stokes $I$ (black) and best-fit profiles obtained with the SIR code and a varying number
of nodes (red $-1 -$, blue $-2-$ and green $-5-$). {\it Bottom panel}: temperature stratification
from the HSRA, VALC and HOLMU (black solid, dotted, and dashed lines, respectively). The color-code 
is the temperature inferred by the application of the SIR inversion code using a varying number of nodes as in the top panel.}
\label{fig:calibration_verification}
\end{figure}

We note that dividing all observed Stokes $I$ spectra by the empirically determined Fourier-filtered curve as described 
in Sect.~\ref{section:observations} can alter the actual shape of the spectral lines. In particular, small ($\approx 1 \%$) 
changes in line-depth and line-width may appear. The same problem would arise if the spectral PSF employed is not accurate. 
Consequently the temperature dependence with optical depth inferred from the inversions (see Sect.~\ref{section:inversion}) 
could be unrealistic. To make sure that this is not the case,  in
this section we perform a series of inversions using the SIR
code \citep[Stokes Inversion based on Response functions; ][]{Cobo_Iniesta_1992} 
to fit the average Stokes $I$ after applying the normalization curve and employing the spectral PSF 
described above. This inversion yields the temperature stratification $T(\log\tau_5)$ of the average quiet Sun, where $\tau_5$ 
indicates the continuum optical depth evaluated at a reference wavelength of 5000 {\AA}. As such, the retrieved $T(\log\tau_5)$ should 
be similar to the temperature stratification of the quiet Sun. The top panel in Fig.~\ref{fig:calibration_verification} 
shows the average quiet Sun Stokes $I$ signal (black) and best-fit profiles obtained with the SIR code with one (red), two (blue) 
and five (green) nodes in temperature. The nodes define the number of optical depth locations $\log(\tau_5)$ where the temperature 
is modified at each iteration step during the inversion. They correspond, therefore, to the number of degrees of freedom allowed during the inversion.
 Different nodes are employed because the normalization procedure described above introduces changes in the intensity profiles that vary across 
wavelength, thereby potentially introducing errors in the retrieved temperature at different optical depths. The bottom panel displays the 
temperature stratification $T(\log\tau_5)$ inferred by the SIR code with these nodes, along with the temperature from three commonly used 
semi-empirical quiet Sun models: the Harvard Smithsonian Reference Atmosphere \citep[HSRA ;][]{hsra1974} (solid black), the Holwerger-M\"uller model \citep[HOLMU ;][]{holweger1974} 
(dashed black), and the VALC model \citep[][]{vernazza1981} (dotted black). As Fig.~\ref{fig:calibration_verification} demonstrates, the inferred temperatures 
resemble that of quiet Sun models, thus indicating that our continuum normalization and spectral PSF are adequate, or at the very least we do not
introduce uncertainties in the temperature larger than those already existing between semi-empirical models of the quiet Sun. Differences for $\log\tau_5 < -2$ can 
be safely ignored since these spectral lines have very little sensitivity above this region (see Section~\ref{section:inversion} 
and Fig.~\ref{fig:response_functions}).\\

\section{Data analysis}
\label{section:analysis}

\subsection{Region of interest}

\begin{figure}
\resizebox{\hsize}{!}{\includegraphics{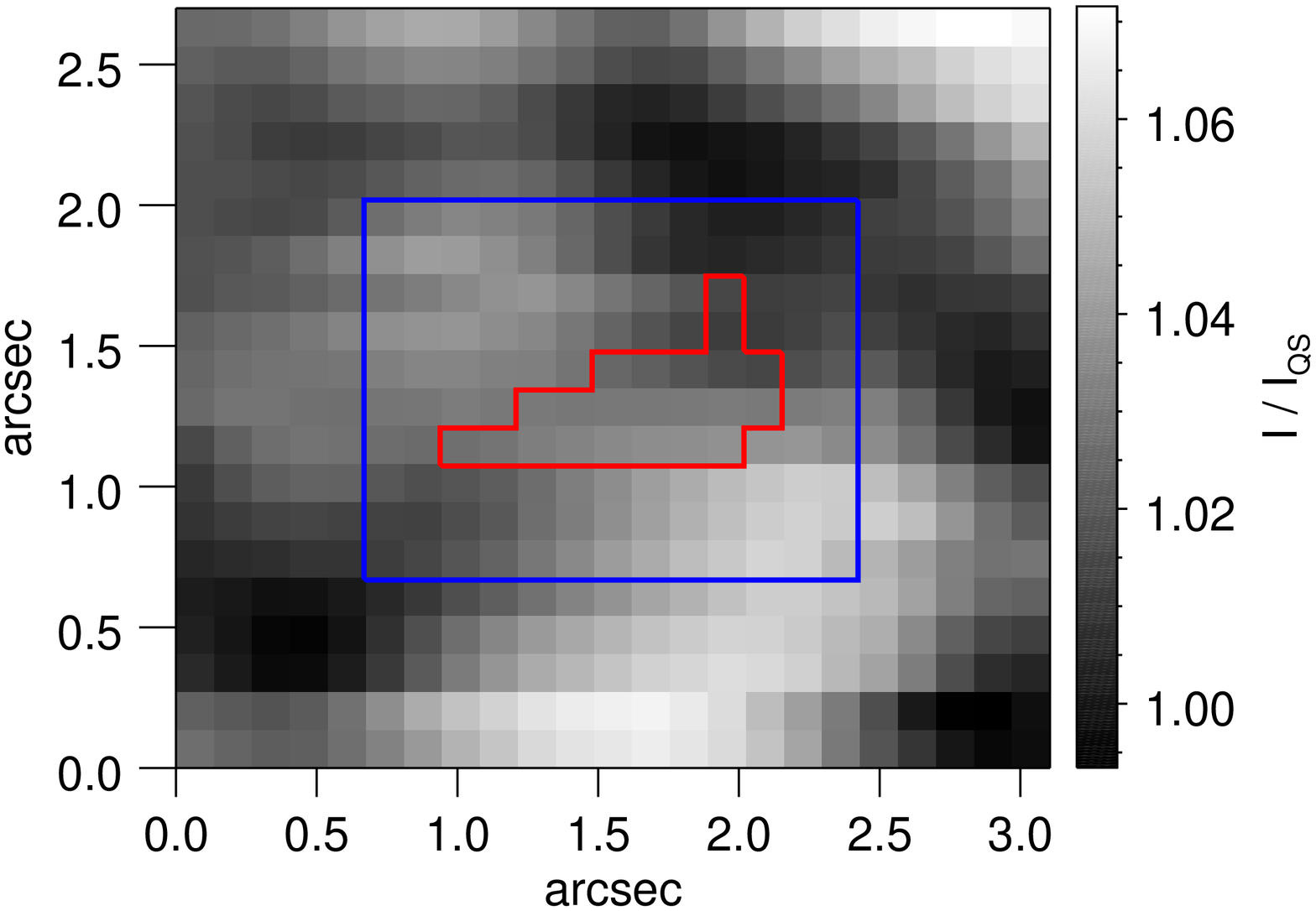}}

\resizebox{\hsize}{!}{\includegraphics{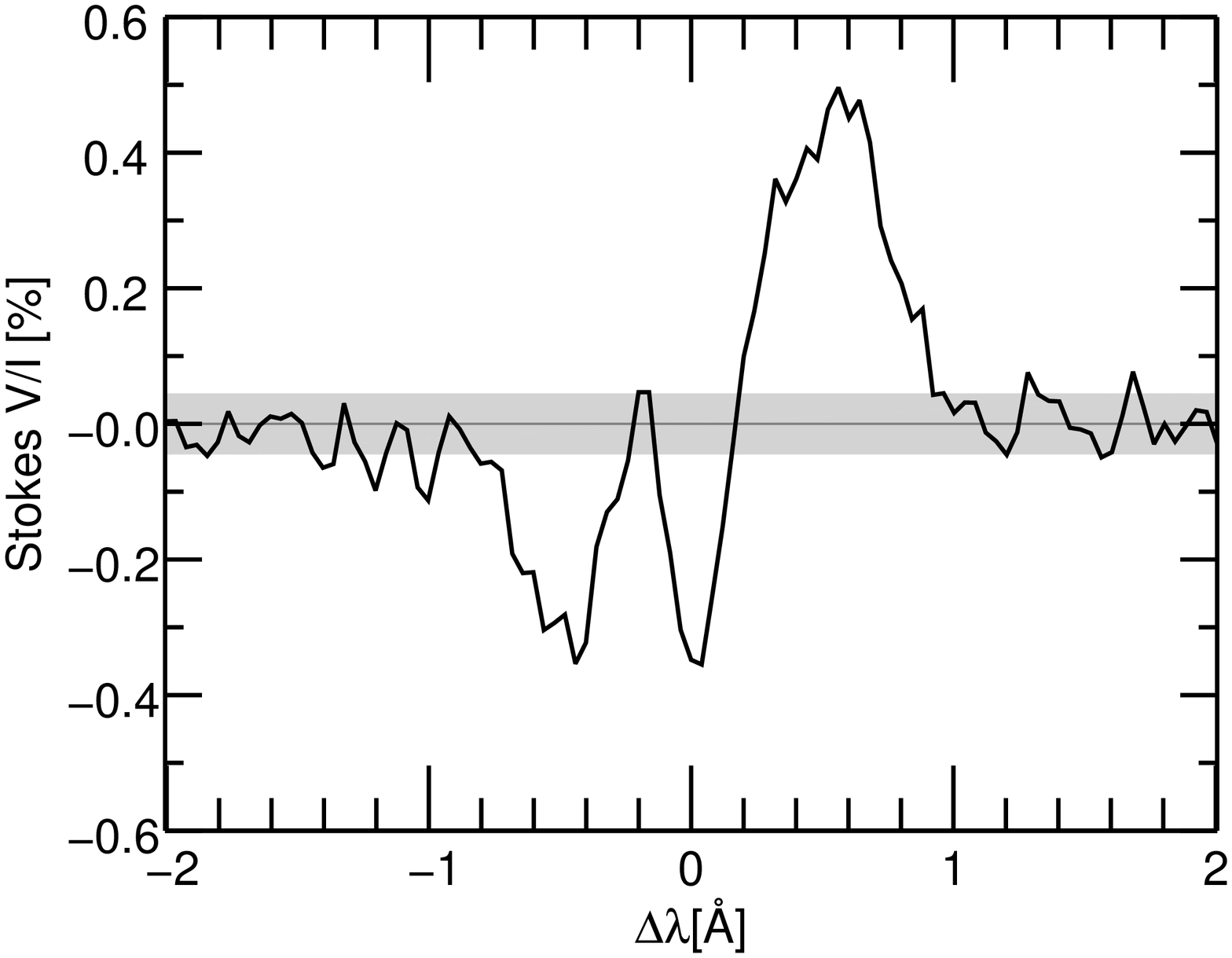}}
\caption{{\it Top panel}: continuum intensity close-up of the analyzed region. See Fig.~\ref{fig:continuum} for the full field-of-view. The red and blue contours 
encircle pixels that were selected for further analysis. {\it Bottom panel}: representative three-lobed circular polarization profile of the 
Fe \textsc{I} 15468 {\AA}. The black line corresponds to a single pixel. The gray area shows the noise level of $4.5\cdot10^{-4}$.}
\label{fig:continuummap_contour}
\end{figure}

We used the Stokes $V$ signal of the Fe \textsc{I} at 15648 {\AA} to identify three-lobed Stokes $V$ profiles. Multi-lobed
circular polarization profiles appear predominantly in this spectral line because it possesses the largest Land\'e factor 
of all the lines included in the observed wavelength range (see Table~\ref{tab:lines}). 
The blue rectangle in Fig.~\ref{fig:continuum} shows the 
position of a large patch containing three-lobed Stokes $V$ profiles. The top panel in Fig.~\ref{fig:continuummap_contour} shows 
a close-up of this region. An example of the observed circular polarization is given on the bottom panel
of this figure. We note that, although the polarization levels are small ($0.4-0.6$ \% in units of the quiet Sun continuum intensity),
in this example the lobes are about ten times above the noise level (indicated by the shaded area on the bottom panel). Of course, three-lobed
Stokes $V$ are not always as clear as in this example. For comparison purposes we provide in the appendix (Figs.~\ref{fig:appendix_first}
through~\ref{fig:appendix_fourth}) a sample of the profiles contained within the blue rectangle. In any case, the signals enclosed by the red
contour in Fig.~\ref{fig:continuummap_contour} (top panel) are such that all three lobes in Stokes $V$ of the Fe \textsc{I} 15648 {\AA} spectral
line are always above three times the noise level. In Section.~\ref{section:inversion} we analyze this region in more detail.\\

\subsection{Spectral analysis}
\label{section:quickanalysis}

Before proceeding with the inversion of Stokes profiles, it is worthwhile learning as much as possible
from the data by performing a qualitative analysis of the observations. This gives us information
about the possible values that the physical parameters might adopt.

\subsubsection*{Magnetic field strength}

In the strong field regime \citep[e.g.,][Eq.~3.55]{Stix_2004} we can calculate the magnetic field strength $B$ with:

\begin{equation}
\lambda - \lambda_0 \approx 4.67 \cdot 10^{-13} g_\text{eff} \lambda^2 B  \;,
\end{equation}

\noindent where $\lambda$ is expressed in {\AA} and $B$ is in Gauss. The above formula gives the line separation 
between the left and right lobes of the Stokes $V$ profile  $\Delta \lambda = 2(\lambda - \lambda_0)$. Therefore:

\begin{equation}
B \approx \frac{\Delta \lambda}{2 \cdot 4.67 \cdot 10^{-13} \cdot \lambda^2 g_\text{eff}} \;.
\label{eq:bfield}
\end{equation}

The separation of the two lobes of the Stokes $V$ profile is roughly 1 {\AA} (see the bottom panel in 
Fig.~\ref{fig:continuummap_contour}), corresponding to a magnetic field strength of about 1500 G. This gives us a 
rough estimation for the magnetic field strength we can expect to be present in the selected region. 
Interestingly, with such large values for the magnetic field, one would expect the amplitude of Stokes $V$
to be much larger than 1 \% and for Stokes $I$ to show some signatures of the Zeeman pattern. However none of
these are observed. It is therefore plausible that only a small portion of the plasma harbors such a strong magnetic field. 
This magnetized region could occupy either a thin vertical layer or a small horizontal and still unresolved fraction of the resolution element.\\

\subsubsection*{Temperature gradient}

The small observed circular polarization amplitude could also be caused by a temperature
stratification varying very slowly with optical depth (i.e., shallow temperature gradient). However, if this were the
case, Stokes $I$ would also be very shallow, which is at odds with the fact that the line-core intensity
in Stokes $I$ is not very different from the line-core intensity of the average quiet Sun. For instance,
the line-core intensity for Fe \textsc{I} 15648 {\AA} is about 0.77 in the selected region (see top-left
panels in Figs.~\ref{fig:pixel_1c} and \ref{fig:pixel_2c}) and 0.74 in the average quiet Sun profile (bottom
panel in Fig.~\ref{fig:continuum_fts_fft} or top panel in Fig.~\ref{fig:calibration_verification}). The discrepancy
between Stokes $I$ and $V$ could be explained if, again, we invoke the presence of unresolved structure
(either vertically or horizontally) within the resolution element, where some region of the solar photosphere
would have a very shallow temperature gradient and strong magnetic field.

\subsubsection*{Inclination of the magnetic field}

As can be seen in Figs.~\ref{fig:pixel_1c} and ~\ref{fig:pixel_2c} and also in Figs.~\ref{fig:appendix_first}-\ref{fig:appendix_fourth},
Stokes $Q$ and $U$ are always below the noise level. In these NIR spectral lines, this 
indicates that the magnetic field is close to being aligned with the observer's line-of-sight \citep{marian2008qs}. 
It is also possible that the field lines are arranged in a way that $Q$ and $U$ cancel themselves out.\\

\subsubsection*{Velocity}

The Stokes $V$ signal is not centered at the same wavelength position as the Stokes $I$ signal (see for instance
Figs.~\ref{fig:pixel_1c} and ~\ref{fig:pixel_2c}). This reinforces the idea of 
a complex atmosphere with vastly different physical properties either stacked up vertically or lying 
next to each other.\\

The separation of the two negative polarity lobes in the blue wing of the Stokes $V$ signal is roughly
500 m{\AA} (see bottom panel in Fig.~\ref{fig:continuummap_contour}). This corresponds to a velocity difference of
about 9 km s$^{-1}$.
While the separation is most likely due to magnetic field strength, it gives an upper limit of possible velocity 
difference in the observed region.

\subsubsection*{Unresolved structure and Stokes $V$ area asymmetry}

We have already gathered evidence pointing towards the existence of variations in the physical parameters
within the resolution element. The question that arises next is whether these variations occur predominantly
along the ray-path (i.e., line-of-sight) or perpendicularly to it. In the former case we speak about variations of the physical
parameters with the optical depth $\tau_5$, whereas in the latter case we refer to horizontal variations
within the resolution element or horizontally unresolved structure. The area asymmetry, $\delta a$,
in the circular polarization is a very useful tool that can be employed to distinguish between
these two limiting cases. It is defined as the integral over wavelength of the Stokes $V$ signal normalized to its total area:

\begin{equation}
\delta a = \frac{\int V(\lambda) {\rm d}\lambda}{\int \|V\|(\lambda) {\rm d}\lambda} \;.
\label{eq:da}
\end{equation}

As demonstrated by \cite{landolfi1996ncp} a sufficient and necessary condition for $\delta a \ne 0$
is the presence of a gradient in the line-of-sight velocity with optical depth (${\rm d} v_{\rm los}/{\rm d} \tau_5 \ne 0$).
The introduction of gradients in other physical quantities such as the magnetic field strength $B$ or the
inclination of the magnetic field with respect to the line-of-sight $\gamma$ increases the absolute value of
$\delta a$ of the observed Stokes $V$ signal \citep{Sanchez_Almeida_1992,solanki1993ncp}.\\

Most of the observed Stokes $V$ signals that we are analyzing (see e.g., bottom panel in Fig.~\ref{fig:continuummap_contour})
have low area asymmetries $\delta a < 0.1$ (with an average absolute value of $\overline{\| \delta a \|} \approx 0.05$).
Usually this is taken as an indication that the three-lobe structure in the circular 
polarization is not due to variations of the velocity and magnetic field with optical 
depth, but instead produced by the presence of horizontally unresolved components within the resolution element.
However, as demonstrated by \cite{Borrero_2004} (see their Fig.~8) the NIR spectral lines employed in this work
show small amounts of area asymmetry even in the presence of large gradients along the line-of-sight in the physical parameters. Because
of this we cannot rule out the existence of such variations. Consequently, we must study these two cases
separately.\\

\section{Stokes inversion set up}
\label{section:inversion}

We used the SIR code \citep{Cobo_Iniesta_1992} to simultaneously invert the Stokes profiles of
the Fe \textsc{I} spectral lines at 15648 {\AA}, 15652 {\AA}, and 15662 {\AA} (see Table~\ref{tab:lines}). 
Details about inversion codes for the radiative transfer equation applied to polarized light can be found in 
the recent review by \cite{Iniesta_Cobo_2016}.
Of particular interest is the concept of \emph{nodes}, which are locations
in the optical depth scale where the derivatives of the Stokes vector with
respect to the physical parameters are evaluated (see Sect.~7.2 of the aforementioned
review). These derivatives are used by the minimization algorithm to perturb the
physical parameters at each iteration step so as to improve the $\chi^2$ merit-function
between the observed and theoretical Stokes parameters \citep[][Eq.~35]{Iniesta_Cobo_2016}.
The total number of nodes during the inversion equals the number of free parameters.\\

During the minimization of the $\chi^2$ merit-function, some Stokes parameters
can be given more priority than others by changing the weights that each of them
have in the calculation of $\chi^2$. In the present work our goal is to 
fit all four measured Stokes profiles reasonably well, with a particular emphasis on capturing 
the three-lobed feature observed in the Fe \textsc{I} 15648 {\AA} line. Therefore Stokes $V$ is 
weighted above the other Stokes parameters ($I$, $Q$ and $U$). More details are given
below.\\

Before proceeding with the inversion of the entire region of interest (see blue square in
the top panel of Fig.~\ref{fig:continuummap_contour}) we tested our inversions on a single
pixel, inside this region, where the three-lobed Stokes $V$ is very prominent, and made sure that
the selected weights produce good fits to the observations. In addition to this, the inversion of each
pixel was repeated 50 times with different random initializations \citep[see][]{Borrero_2017}.
The result with the best $\chi^2$ from those 50 inversions was then selected for that particular pixel.
This was done in order to prevent the Levenberg-Marquardt inversion algorithm \citep{Press_1986} from
falling into any local minimum in the $\chi^2$-hypersurface. We note that the highest atmospheric
layer in our atmospheric models is $\log\tau_5=-3$ so as to avoid having nodes in regions of low
sensitivity to the various physical parameters (see Section~\ref{section:results} for more details).\\

In order to fit the multi-lobe structure of the circular polarization signals,
variations in the velocity and magnetic field must be included and permitted during 
the inversion. Unfortunately (see Sect.~\ref{section:quickanalysis}) the study of the area 
asymmetry did not allow us to disentangle the presence of vertical or
horizontal variations. Therefore we study both possibilities independently.\\

\subsection{One-component inversion}
\label{section:one_component}

Here we assume that the entire resolution element is filled with one single
component where the physical parameters vary with optical depth. During the inversion
we found that the SIR code could not fit all four Stokes parameters simultaneously 
while reproducing the three-lobed pattern in Stokes $V$. This happened even when increasing the
weight given to Stokes $V$. Because of this we split the inversion in two different stages.
In the first stage only Stokes $I$ and $V$ were inverted: weights to $Q$ and $U$ were set to zero
and Stokes $V$ was given 20 times more weight than Stokes $I$. The temperature $T(\tau_5)$, magnetic field strength $B(\tau_5)$,
inclination of the magnetic field with respect to the observer's line-of-sight $\gamma(\tau_5)$,
and line-of-sight velocity $v_{\rm los}(\tau_5)$ were allowed to change in three nodes each. 
Microturbulence was allowed to change with one node. The azimuth in this first step was not allowed to change 
and was kept at zero.\\

Subsequently, a second inversion stage was performed using the results from
the previous stage for initialization and now including $Q$ and $U$, which were given the same weight as Stokes $I$
(20 times less than Stokes $V$). All physical parameters were kept the same as the results
from the first stage, except for the azimuth of the magnetic field in the plane perpendicular to 
the line-of-sight $\phi(\tau_5)$ which was allowed to change in two nodes.\\

\begin{figure}
\resizebox{\hsize}{!}{\includegraphics{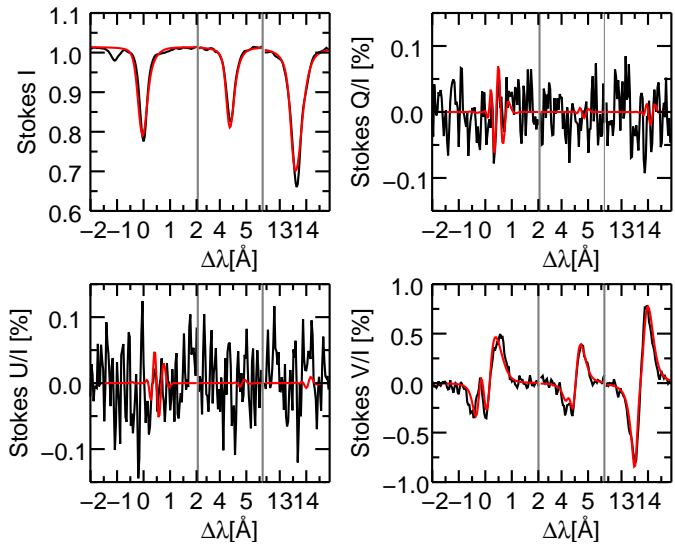}}
\caption{Representative pixel for one-component fit. The black curve shows the observed Stokes profiles: $I$ (top-left),
$Q$ (top-right), $U$ (bottom-left), and $V$ (bottom-right). The red curve shows the best-fit profiles obtained from
the inversion (i.e., smallest $\chi^2$).}
\label{fig:pixel_1c}
\end{figure}

Combining both stages the total number of free parameters employed was 15. This procedure allowed 
us to fit reasonably well the polarized Stokes parameters ($Q$, $U$, and $V$) including the three-lobed 
structure in the circular polarization (see Fig.~\ref{fig:pixel_1c}). Owing to the smaller weight given 
to Stokes $I$ during the inversion, the fits in Stokes $I$ are of lesser quality than those of Stokes $V$. Moreover, 
as can be seen in this figure, the depth of Stokes $I$ for the Fe \textsc{I} spectral line at 15662 {\AA} 
(i.e., third line from the right in Figs.~\ref{fig:pixel_1c},~\ref{fig:pixel_2c}) is particularly poorly reproduced. 
This could be due to the oscillator strength employed for this line not being as accurate as the others 
(see Table~\ref{tab:lines}). We note that the presence of telluric and molecular blends, interference fringes, assumption
of LS coupling, and so on, make the intensity profiles of these lines in the NIR notoriously difficult to fit
\footnote{Poor fits in Stokes $I$ can also seen in Figs.~3-4 in \citep{cabrera2008}, Fig.~1 in \citep{bellot2000},
Fig.~2 in \citep{Toro_Iniesta_2001}, Fig.~6 in \citep{Borrero_2016}. Some of these works obtained poor fits 
in spite of employing a depth-dependent (i.e., more than one node) microturbulent velocity.}\\

\subsection{Two-component inversion}
\label{section:two_component}

In the two-component inversion we assume that the resolution element is composed of two
distinct atmospheric components that are interlaced horizontally. The first one covers a fraction
of the total area of the resolution element $\alpha$. The area occupied by the second
component would therefore be $(1-\alpha)$. 
In order to clearly separate the inversions in this section from the previous ones we take the magnetic
and kinematic parameters to be constant with optical depth in both components. In this fashion the three-lobed
structure of the observed circular polarization profiles (see bottom panel in 
Fig.~\ref{fig:continuummap_contour}) is completely ascribed to the horizontal variations
(i.e., differences between the first and second components) of the physical parameters
as opposed to the vertical variations of the one-component inversions (Sect.~\ref{section:one_component}).
We note that the temperature in both components must be dependent on $\tau_5$  otherwise
Planck's function would be the same at all photospheric layers and there would be no
spectral features (i.e., no absorption in Stokes $I$ and only a flat continuum).\\

During the inversion, the temperature $T$ for each component was allowed to change in three nodes, 
while for $B$, $v_{\rm los}$ and microturbulence we employed only one node (i.e., constant values with $\tau_5$) per component.
Together with the filling factor $\alpha$ this adds up to a total of 13 
free parameters during the inversion. For reasons that will become apparent 
later, the magnetic field inclination was fixed and not allowed to change during the inversion to $\gamma=0^{\circ}$ 
for the first component (constant with $\tau_5$) and $\gamma=180^{\circ}$ for the second component.
Just as before, the weight of Stokes $V$ was 20 times larger than for Stokes $I$. Since the
inclination of the magnetic field is such that the magnetic field is always aligned with the 
observer's line-of-sight (parallel for $\gamma=0^{\circ}$ and antiparallel for $\gamma=180^{\circ}$)
this set-up does not produce any linear polarization and therefore the weights for Stokes $Q$
and $U$ were set to zero. Unlike the one-component inversion in the previous section
here the inversion was performed in a single stage.\\

\begin{figure}
\resizebox{\hsize}{!}{\includegraphics{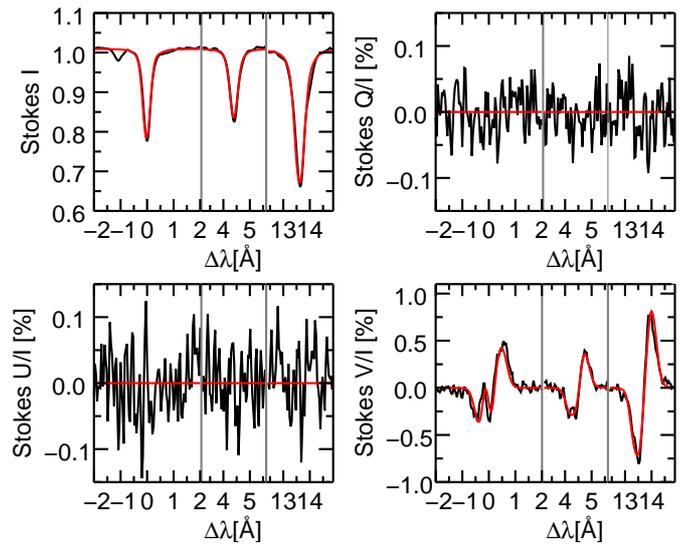}}
\caption{As in Fig.~\ref{fig:pixel_1c} but for the two-component inversion.}
\label{fig:pixel_2c}
\end{figure}

The fit to the observed Stokes profiles achieved with the aforementioned configuration
is of similar quality to the one-component inversion. An example can be seen in Fig.~\ref{fig:pixel_2c}.
Owing to the fact that we have twice as many free parameters in the temperature than in the one-component
inversion, the quality of the fits to Stokes $I$ has increased.\\

\section{Stokes inversion results}
\label{section:results}

\begin{figure}
\resizebox{\hsize}{!}{\includegraphics{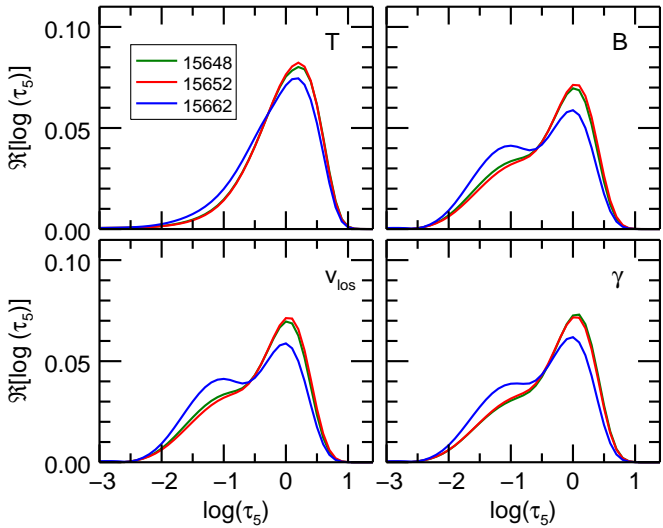}}
\caption{Response functions integrated over wavelength, added up for all four Stokes parameters
and plotted as a function of the logarithm of the optical depth $\log\tau_5$. {\it Top-left}: response 
to the temperature $T$. {\it Top-right}: response to the line-of-sight velocity $v_{\rm los}$. {\it Bottom-left}: response 
to the magnetic field strength $B$. {\it Bottom-right}: response to the inclination of the magnetic field with 
respect to the observer's line-of-sight $\gamma$. Different colors correspond to different spectral lines.
{\it Green}: 15648 {\AA}, {\it red}: 15652 {\AA}, {\it blue}: 15662 {\AA}. The response functions are calculated with 
the atmosphere resulting from the two-component inversion shown in Fig.~\ref{fig:pixel_2c}.
\label{fig:response_functions}}
\end{figure}

Figure~\ref{fig:response_functions} displays a measure of the sensitivity to the physical parameters 
for the spectral lines in Table~\ref{tab:lines}. It was obtained by adding up the individual unsigned response functions of the four Stokes 
parameters using the weights employed 
during the inversion (see Sect.~\ref{section:inversion}), integrating in wavelength and normalizing the resulting curve to have an area equal to one. The model employed to calculate 
these response functions is the average model from the two-component inversion 
(see Sects.~\ref{section:one_component} and ~\ref{section:one_component_result}). The resulting functions show 
that the NIR lines discussed in this paper form deep in the solar photosphere, in particular 
when compared to the commonly used Fe \textsc{I} spectral lines in the visible \citep[see also][]{Borrero_2016}.
While our models reach up to $\log\tau_5=-3$ (see Sect.~\ref{section:inversion}), 
most of the sensitivity of the spectral lines used in this work comes from a thin photospheric layer 
located between $\log\tau_5 \in [-2,0.5]$. The temperature is sensitive to slightly deeper layers compared 
to the other atmospheric parameters. Outside this range the response functions
are too small and therefore the uncertainty in the inference of the physical parameters increases.
Because of this, in the following we  
only discuss the inversion results obtained in the range 
$\log\tau_5 \in [-2,0.5]$.\\

\subsection{One-component result}
\label{section:one_component_result}

\begin{figure}
\resizebox{\hsize}{!}{\includegraphics{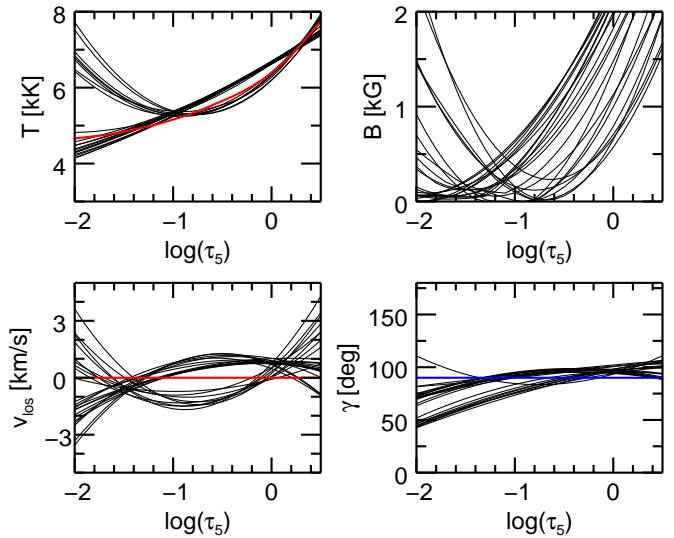}}
\caption{Results as a function of the logarithm of the optical depth ($\log\tau_5$) for the one 
component inversion: temperature $T$ (top-left), magnetic field strength $B$ (top-right), line-of-sight
velocity $v_{\rm los}$ (bottom-left), inclination of the magnetic field with respect to the
observer's line-of-sight $\gamma$ (bottom-right). Individual black curves correspond to the inversion
of individual pixels. The red curve on the top-left panels shows the temperature stratification
from the HSRA model. Red/blue curves on the lower panels indicate $v_{\rm los}=0$ (bottom-left)
and $\gamma=90^{\circ}$ (i.e., horizontal magnetic field; bottom-right).}
\label{fig:atm_height_1c}
\end{figure}

The inferred physical parameters as a function of optical depth from the one
component inversion (Sect.~\ref{section:one_component}) are presented in Fig.~\ref{fig:atm_height_1c}. The results
for the individual pixels within the region of interest within the red contour in Fig.~\ref{fig:continuummap_contour}
(top panel) are indicated by the black lines. We find that the temperature in the range
$\log\tau_5 \in [-2,0.5]$ closely follows that of the HSRA atmosphere (red line
in the top-left panel). The magnetic field (top-right panel) quickly decreases 
from kilo-Gauss values in the deep photosphere ($\log\tau_5 \approx 0$) to almost zero
approximately 150-200 km higher up ($\log\tau_5 \approx -1.5$). The photospheric layer
with large magnetic fields possesses redshifted velocities that turn into
blueshifts as the magnetic field decreases higher up. In both cases, the absolute velocities
are about 1-2 km s$^{-1}$ (bottom-left panel). It must be borne in mind 
that a change of sign in $v_{\rm los}$ can be due solely to projection effects (i.e., observer's
line-of-sight).\\

The magnetic field is highly inclined with respect to the observer ($\gamma \approx 90^{\circ}$; bottom-right
panel) at all optical depths. At closer inspection we see that it points slightly away from the observer in the strong
field region ($\gamma(\log\tau_5 \approx 0) > 90^{\circ}$) and slightly towards the observer
higher up in the weak-field region ($\gamma(\log\tau_5 \approx -1.5) < 90^{\circ}$).
The change of polarity (with respect to the observer) in the inclination of the magnetic field,
as well as the change from blueshifted to redshifted velocities, are key features in reproducing
three-lobed circular polarization profiles in the analyzed NIR spectral lines. They have
previously been reported in other solar regions, such as sunspot penumbrae \citep[see Fig.~3 in][]{Borrero_2004}.
To demonstrate the need for a polarity change we have conducted two experiments. In the first one 
we made $\gamma=90^{\circ}$ whenever $\gamma > 90^{\circ}$ in Fig.~\ref{fig:atm_height_1c} while keeping all other physical 
parameters the same, and using the resulting atmospheric model to create synthetic profiles of the observed spectral lines.
In the second one we did the opposite, that is, we kept $\gamma=90^{\circ}$ whenever $\gamma < 90^{\circ}$ while 
keeping all other physical parameters the same. In both cases, the polarity change is effectively removed. Interestingly,
in both instances, the third lobe in Stokes $V$ of the Fe {\sc I} 15648.5 {\AA}~ immediately disappears, thereby proving that the change 
in polarity is a necessary feature.\\

\begin{figure}
\resizebox{\hsize}{!}{\includegraphics{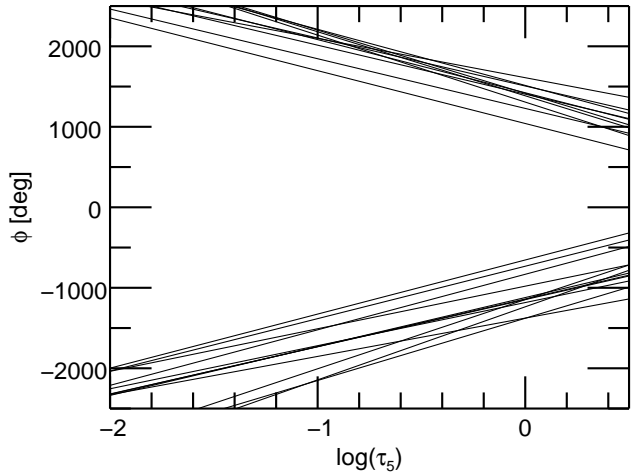}}
\caption{Azimuth of the magnetic field in the plane perpendicular to the observer's
line of sight $\phi$ as a function of the logarithm of the optical depth $\log\tau_5$
for the one-component inversion. Individual black curves correspond to the inversion
of individual pixels. The average gradient is 700$^{\circ}$ per optical depth scale decade, indicating 
multiple rotations of the magnetic field vector within the photosphere.}
\label{fig:phi_height_1c}
\end{figure}

It is important to mention here that, at each iteration step, the SIR inversion code performs a smooth interpolation 
in $\tau_5$ of the physical parameters between the values at the nodes. As a consequence, if 
the code finds it necessary to change from $\gamma < 90^{\circ}$ to $\gamma > 90^{\circ}$ (i.e., polarity change) 
to fit the multi-lobe structure of Stokes $V$, it follows then that $\gamma \approx 90^{\circ}$ somewhere in the
optical depth region where the line is formed. Such a highly inclined magnetic field will produce a strong linear polarization signal. However,
the observed $Q$ and $U$ profiles are below the noise level (see black lines in Fig.~\ref{fig:pixel_1c}). 
The inversion code is therefore compelled to reduce the amount of linear polarization by including large changes
with optical depth in the azimuth of the magnetic field $\phi$. These large variations are displayed in 
Fig.~\ref{fig:phi_height_1c}. Here we can see that $\phi$ changes by about 2000 degrees between
$\log\tau_5 \in [-2,0.5]$. If real, this would imply that over a vertical distance of 200-300 km, the magnetic field
performs about four to five full rotations on the plane perpendicular to the observer's lines-of-sight. 

\subsection{Two-component result}
\label{section:two_component_result}

\begin{figure}
\resizebox{\hsize}{!}{\includegraphics{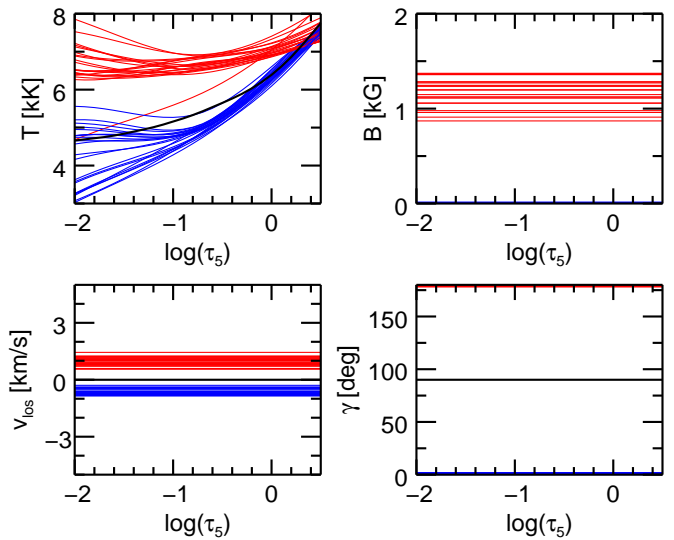}}
\caption{As in Fig.~\ref{fig:atm_height_1c} but for the two-component
inversion. Red and blue colors represent the first 
and second components, respectively . Each curve corresponds to a single pixel.}
\label{fig:atm_height_2c}
\end{figure}

Figure~\ref{fig:atm_height_2c} shows the inferred physical parameters as a function
of the optical depth from the two-component inversion
(Sect.~\ref{section:two_component}). Red curves represent the results for each of the
inverted pixels in the region with the red contour in Fig.~\ref{fig:continuummap_contour} (Top panel)
for the first of the two components. This first component features a hot plasma with a shallow temperature
gradient (top-left panel), a strong magnetic field ($B \gtrsim 1000$; top-right panel), and 
redshifted velocities of about 1.5 km s$^{-1}$ (bottom-left panel). Likewise, the blue curves 
represent the physical parameters for the second component. Here, the temperature is colder, 
at $\log\tau_5=0$, by about 1000 K than in the first component. This temperature difference 
increases towards higher layers and reaches about 3000 K at $\log\tau_5=-2$. The magnetic field 
in this component is much weaker and very close to zero ($B \approx 10$) and the line-of-sight velocities are
blueshifted by about 0.6 km s$^{-1}$. By construction (see Sect.~\ref{section:two_component}) the magnetic field 
is always vertical in both components, but it points away from ($\gamma=180^{\circ}$) and towards ($\gamma=0^{\circ}$) 
the observer in the hot and cold components, respectively. We note that 
it was not possible to fit the third lobe in Stokes $V$ if both components had the
same polarity\footnote{A single-polarity atmosphere also failed to reproduce the three-lobe Stokes $V$
in the one-component inversion (see Sect.~\ref{section:one_component_result})}. An azimuth of the magnetic field was not determined since the  
magnetic field vector is parallel to the line-of-sight.
The filling factor of the first component (i.e., strong field component) is within $\alpha \in [0.2,0.4]$ for most pixels.\\

\begin{figure}
\resizebox{\hsize}{!}{\includegraphics{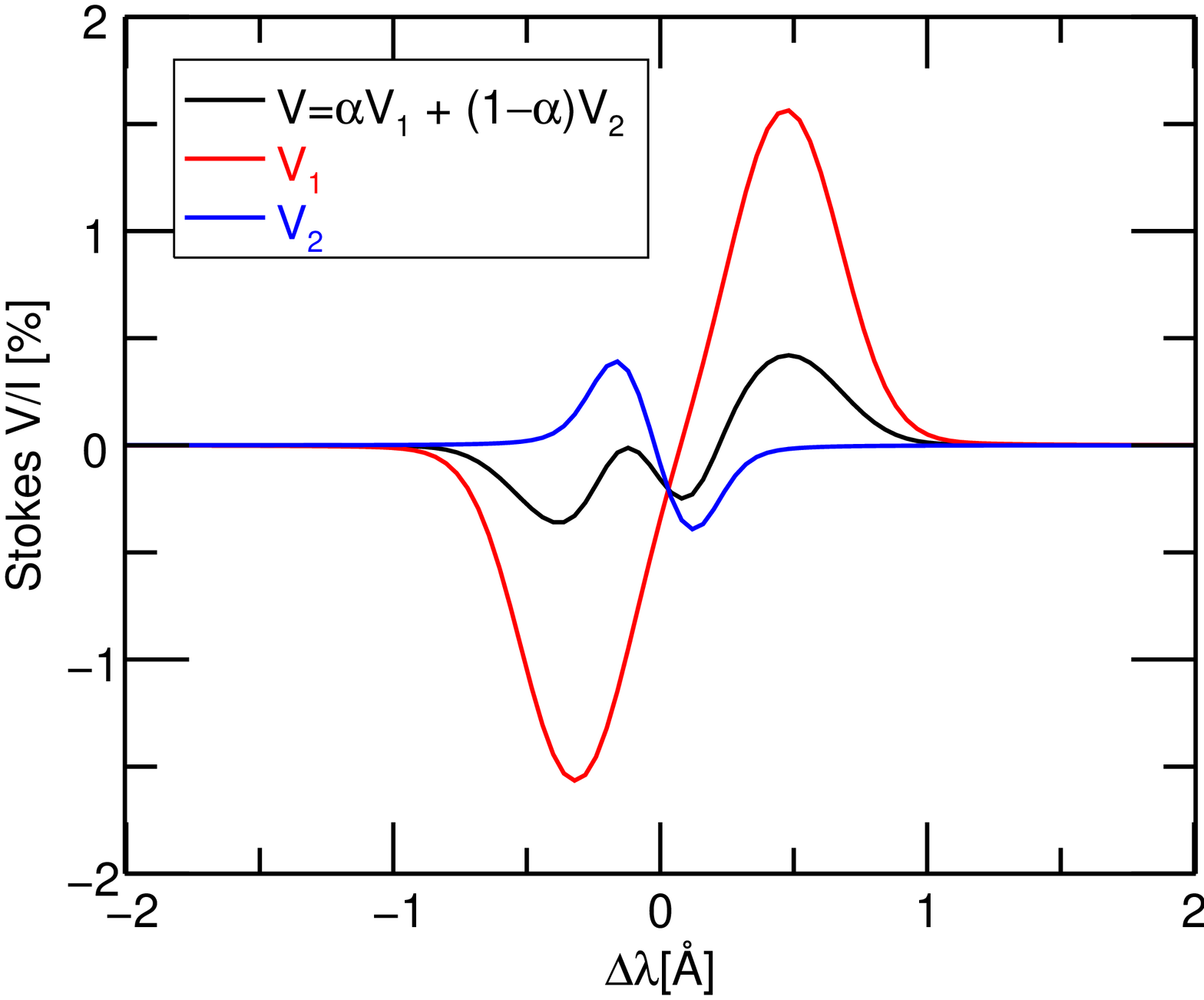}} 

\resizebox{\hsize}{!}{\includegraphics{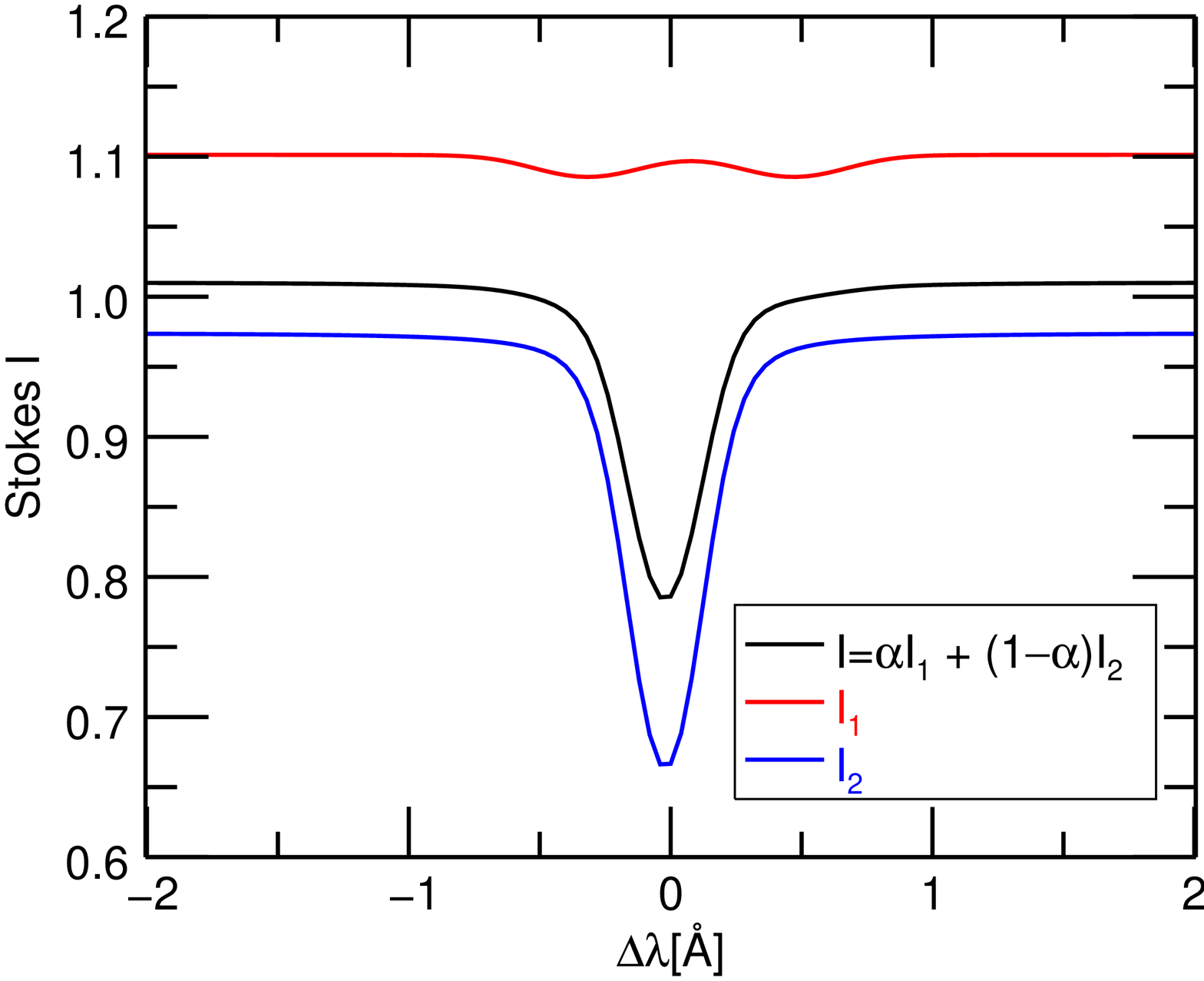}}
\caption{Comparison of synthetic Stokes $V$ (upper panel) and $I$ (lower panel) profiles of Fe I 15648 {\AA}~
in a two-component atmosphere. The black curve is the inversion result presented in Fig. ~\ref{fig:pixel_2c}.
The red (blue) profile is caused by only the first (second) component, respectively. The respective magnetic 
field strengths are $B = 1147~\rm{G}$  and $B=9~\rm{G}$ and the filling factor of the component 
with the strong magnetic field is $\alpha = 0.27$. We note that both components produce circular polarization 
signals of similar amplitudes in spite of the very different magnetic fields. The final
continuum intensity (lower panel) is about 1.0 (normalized to the quiet Sun value) in spite of the first component featuring
very high temperatures (see red lines in the upper-left panel of Fig.~\ref{fig:atm_height_2c}).}
\label{fig:two_components_synthesis}
\end{figure}

One could be tempted to argue that the magnetic field in the second component, being so low,
is indeed zero or at least negligible. However, as we demonstrate in Fig.~\ref{fig:two_components_synthesis}
(upper panel) a $B \ge 0$ in the second component is absolutely necessary if the three-lobed structure
in the circular polarization is to be reproduced. This figure shows the Stokes $V$ profiles
of the spectral line Fe \textsc{I} 15648 {\AA} created by the first and second components in
red and blue colors, respectively. The black line corresponds to the final Stokes $V$ resulting
from combining the previous two with a filling factor for the first component of $\alpha=0.27$. 
Due to its larger
magnetic field, the level of circular polarization created by the second component is larger than that of the first one, and in spite of its shallower temperature decrease
with optical depth. However, the amount of Stokes $V$ generated by the second component becomes
comparable to that of the first component once we consider that the latter occupies a larger fraction of the resolution element (0.73
vs. 0.27). When comparing the Stokes $I$ profiles emerging from each of the two components individually (lower panel
in Fig.~\ref{fig:two_components_synthesis}) we see that the first component, being much hotter at $\tau_5=1$ than
the average quiet Sun, produces a continuum intensity of 1.1 (red), whereas the second component, having a similar temperature at $\tau_5=1$ to that of the average quiet Sun, 
produces also a similar continuum intensity of 0.97 (blue). Once both components are added, considering a filling factor
of $\alpha=0.27$ for the first component, the resulting continuum intensity is very close to 1.0 (as observed in Figs.~\ref{fig:pixel_1c}
and \ref{fig:pixel_2c}).\\

\section{Discussion}
\label{section:discussion}

Despite the very different geometrical configurations employed (Sect.~\ref{section:inversion}), 
the one- and two-component inversions yield physical parameters that are largely consistent with each other. 
In both cases, we detect, within the resolution element, regions of the photosphere 
harboring a strong magnetic field ($B \ge 1000$ G), that point away from the observer ($\gamma > 90^{\circ}$), 
and where the velocity is redshifted ($v_{\rm los} > 0$). Along with this, we find other regions within the 
resolution element with a much weaker magnetic field ($B \approx 10$ G), that points towards 
the observer ($\gamma < 90^{\circ}$), and where the velocity is blueshifted ($v_{\rm los} < 0$). 
In particular, the physical parameters at $\log\tau_5 \approx 0$ inferred from the one-component 
inversions (Fig.~\ref{fig:atm_height_1c}) are very similar to those of the first component in 
the two-component inversion (red curves in Fig.~\ref{fig:atm_height_2c}). Likewise, the physical 
parameters at $\log\tau_5 \approx -1.5$ inferred from the one-component 
inversions (Fig.~\ref{fig:atm_height_1c}) are very similar to those of the second component in the two-component 
inversion (blue curves in Fig.~\ref{fig:atm_height_2c}). \\

It is not possible at this stage to say whether these two regions are interlaced horizontally 
(i.e., lying next to each other) or vertically (i.e., lying one on top of the other). The latter
case is represented by the one-component inversion, whereas the former case is represented by
the two-component inversion. On the one
hand we cannot employ the quality of the fits in order to decide which type of inversion is to be preferred
because both reproduce reasonably well the observed Stokes profiles (cf. Fig.~\ref{fig:pixel_1c}
and Fig.~\ref{fig:pixel_2c}; see also Fig.~\ref{fig:appendix_first} through Fig.~\ref{fig:appendix_fourth}).
On the other hand, the area asymmetry in these NIR spectral lines (Table~\ref{tab:lines}) is
small even in the presence of very strong gradients along the line-of-sight in the magnetic and kinematic
parameters \cite{Borrero_2004}, and therefore cannot be invoked to decide which inversion is to be preferred 
(see Sect.~\ref{section:quickanalysis}). More complex inversion approaches, in particular multi-component inversions with gradients,
could also be used to reproduce the observed signal. Since this would significantly increase the number of free parameters, we limit our analysis to 
inversions with the least possible number of degrees of freedom that still fit the Stokes $V$ profile.\\

An alternative approach to decide between the one- and two-component inversion is to look at the realism
(or lack of) of the inferred physical parameters. As mentioned in Sect.~\ref{section:one_component_result}, the one-component
inversion leads to the magnetic field performing several rotations (see Fig.~\ref{fig:phi_height_1c}) in the 
plane perpendicular to the observer's line-of-sight along a vertical distance of 200-300 km.
We cannot readily regard this result as unphysical because similar configurations have been previously reported 
for the velocity \citep{bonet2010imax,wedemeyer2012vortex}. In particular, the latter authors find that, in 
three-dimensional MHD simulations performed with the CO$^5$BOLD code, the velocity vector can perform many rotations 
over such a short vertical distance (see their Fig.~3). In our observations we find that the magnetic field could 
also behave in this fashion. On the other hand, these rotations could be simply an artifact of 
the SIR inversion code attempting to fit three-lobed Stokes $V$ profiles while keeping $Q$ and $U$ close 
to zero using smooth variations with $\tau_5$ in the physical parameters. As discussed in 
Sect.~\ref{section:one_component_result} this forces the code to find a solution where the magnetic field changes 
polarity ($\gamma<90^{\circ} \rightarrow \gamma>90^{\circ}$), thus becoming horizontal with respect to the observer
in the middle of the line-forming region ($\gamma(\log\tau_5 \approx -1 )=90^{\circ}$). The large variations
in the azimuthal angle $\phi$ appear then as a consequence of the requirement to make the linear
polarization signals vanish in spite of the horizontal magnetic field. It must be borne in mind that it might be possible to fit the profiles if we 
include a discontinuity in $\gamma$ at around $\log\tau_5 = -1$, with $\gamma = 0^{\circ}$ above and $\gamma = 180^{\circ}$ below. 
Such a configuration would still feature a polarity change, thereby fitting the three-lobed Stokes $V$ profiles, but with a magnetic 
field that is always aligned with the observer's line-of-sight (i.e., either parallel or antiparallel to it) which produces, as observed, 
no linear polarization. In this case there would be no need for any variations of the azimuthal angle $\phi$ with optical depth.\\

Some results from the two-component inversion could also be questioned.
In particular some doubts could be cast on the temperature stratification $T(\log\tau_5)$ of the strong field component
(see red curves on the top-left panel of Fig.~\ref{fig:atm_height_2c}). Such an atmospheric component is hotter than the quiet Sun (black curves in this figure shows the HSRA model)
by 1000-3000 K, and it decays very slowly towards
higher photospheric layers. While seemingly unrealistic, similar temperature stratifications have previously been reported
in unipolar magnetic elements in the quiet Sun \citep{lagg2010imax}. Moreover, regions with multi-lobe Stokes $V$
profiles in the Fe \textsc{I} spectral lines at 6300 {\AA} were analyzed by \cite{Quintero_Noda_2014b} who also found
enhanced temperatures, kilo-Gauss magnetic fields, and large redshifted line-of-sight velocities. These properties are 
somewhat similar to the ones inferred for the strong field component in our two-component inversion. Therefore, the
inferred temperatures cannot be considered as unrealistic and cannot be used to rule out the results from the
two-component inversion.\\

\section{Conclusions and future work}
\label{section:conclusions}

We have analyzed high-quality (i.e., polarimetric noise below $5\times 10^{-4}$ in units of the quiet Sun intensity) and high-spatial-resolution ($\approx 0.4"$) spectropolarimetric data (full Stokes vector) from the quiet Sun recorded with the GRIS instrument 
attached to the 1.5-m German solar telescope GREGOR. The observed spectral region includes three Fe \textsc{I} lines 
located in the near-infrared (15650 {\AA}) that are highly sensitive to the
magnetic field. We focus on a small region ($\approx 0.5"$ in diameter) in the quiet Sun where the circular polarization 
profiles of one of those lines presents three lobes. This feature is indicative of variations in the physical parameters, 
in particular in the magnetic field and line-of-sight velocity, within the resolution element. To investigate whether those 
variations occur predominantly along the line-of-sight or perpendicular to it we have performed two kinds of inversion of the 
observed profiles. In the first one we assume that the resolution element is occupied by one magnetic atmosphere where the 
kinematic and magnetic parameters vary with optical depth (i.e., line-of-sight). In the second one we assume the resolution 
element to be occupied by two magnetic atmospheres where the kinematic and magnetic parameters are constant with optical depth.\\

The analysis of the data using both inversion set-ups reveals that the three lobes seen in Stokes $V$ are produced by the presence of
magnetic fields of opposite polarity. One of them possesses a strong magnetic field
($B \ge 1000$ G) and harbors plasma velocities directed away from the observer, whereas the other one possesses a much weaker field
($B \approx 10$ G) where the velocity is directed towards the observer. In spite of these similarities, each kind of inversion
also presents some peculiarities.\\

The one-component inversion fits the observed profiles by invoking large variations in the azimuth of the magnetic field
(i.e., rotations of the magnetic field vector in the plane perpendicular to the observer's line-of-sight). Such swirling 
or tornado-like configuration has already been found in the velocity \citep{bonet2010imax,wedemeyer2012vortex}. In our work 
we find that the magnetic field could behave similarly. We emphasize that it could also be possible
to fit the observed profiles without such large variations in the azimuth of the magnetic field if we consider 
discontinuities in the inclination of the magnetic field with respect to the observer's line-of-sight.\\

The two-component inversion infers a temperature stratification $T(\log\tau_5)$ in the region with a strong magnetic field that is (much) hotter than the average quiet Sun by 1000-3000
K. High temperatures have been found
in magnetic elements \citep[network, ][]{lagg2010imax}, in strongly magnetized upflow regions inside granules \citep{Borrero_2013}
and downflow regions \citep{Quintero_Noda_2014b}.\\

Therefore, neither the physical realism nor the goodness of the fits to the observations can be invoked to give preference
to one inversion set-up over the other. To answer this question it would be desirable to employ inversion codes for the
radiative transfer equation that allow for discontinuities in the physical parameters 
\citep{louis2009,Sainz_Dalda_2012,marian2012sirjump,shalout2015sirjump}. It would also be appropriate to study whether the Stokes profiles emerging from numerical simulations show a similar
behavior, so as to find out the stratification of the physical parameters in those simulations and compare them to
the inferred stratifications from the observations \citep{Danilovic_2015}.\\
To resolve the ambiguity discussed in this work, studies of multiple events, possibly with higher spatial resolution will be necessary. 
The next generation of large solar telescopes, such as DKIST \citep{Rimmele_2015}, will provide the high-resolution data needed to advance the study of small scale, complex magnetic features.

\begin{acknowledgement}
The 1.5-m GREGOR solar telescope was built by a German consortium under
the leadership of the Kiepenheuer-Institut f\"ur Sonnenphysik in Freiburg with
the Leibniz-Institut f\"ur Astrophysik Potsdam, the Institut f\"ur Astrophysik G\"ottingen,
and the Max-Planck-Institut f\"ur Sonnensystemforschung in G\"ottingen as
partners, and with contributions by the Instituto de Astrof\'isica de Canarias and
the Astronomical Institute of the Academy of Sciences of the Czech Republic.
\end{acknowledgement}

\bibliography{32267}{}

\begin{thebibliography}{71}
\expandafter\ifx\csname natexlab\endcsname\relax\def\natexlab#1{#1}\fi

\bibitem[{{Anstee} \& {O'Mara}(1995)}]{abo1}
{Anstee}, S.~D. \& {O'Mara}, B.~J. 1995, \mnras, 276, 859

\bibitem[{{Auer} \& {Heasley}(1978)}]{auer78ncp}
{Auer}, L.~H. \& {Heasley}, J.~N. 1978, \aap, 64, 67

\bibitem[{{Barklem} \& {O'Mara}(1997)}]{abo2}
{Barklem}, P.~S. \& {O'Mara}, B.~J. 1997, \mnras, 290, 102

\bibitem[{{Barklem} {et~al.}(1998){Barklem}, {O'Mara}, \& {Ross}}]{abo3}
{Barklem}, P.~S., {O'Mara}, B.~J., \& {Ross}, J.~E. 1998, \mnras, 296, 1057

\bibitem[{{Bellot Rubio} {et~al.}(2004){Bellot Rubio}, {Balthasar}, \&
  {Collados}}]{Bellot_Rubio_2004}
{Bellot Rubio}, L.~R., {Balthasar}, H., \& {Collados}, M. 2004, \aap, 427, 319

\bibitem[{{Bellot Rubio} {et~al.}(2000{\natexlab{a}}){Bellot Rubio},
  {Collados}, {Ruiz Cobo}, \& {Rodr{\'{\i}}guez Hidalgo}}]{bellot2000}
{Bellot Rubio}, L.~R., {Collados}, M., {Ruiz Cobo}, B., \& {Rodr{\'{\i}}guez
  Hidalgo}, I. 2000{\natexlab{a}}, \apj, 534, 989

\bibitem[{{Bellot Rubio} {et~al.}(2001){Bellot Rubio}, {Rodr{\'{\i}}guez
  Hidalgo}, {Collados}, {Khomenko}, \& {Ruiz Cobo}}]{Bellot_Rubio_2001}
{Bellot Rubio}, L.~R., {Rodr{\'{\i}}guez Hidalgo}, I., {Collados}, M.,
  {Khomenko}, E., \& {Ruiz Cobo}, B. 2001, \apj, 560, 1010

\bibitem[{{Bellot Rubio} {et~al.}(2000{\natexlab{b}}){Bellot Rubio}, {Ruiz
  Cobo}, \& {Collados}}]{bellot2000ncp}
{Bellot Rubio}, L.~R., {Ruiz Cobo}, B., \& {Collados}, M. 2000{\natexlab{b}},
  \apj, 535, 489

\bibitem[{{Berkefeld } {et~al.}(2012){Berkefeld }, {Schmidt}, {Soltau}, {von
  der L{\"u}he}, \& {Heidecke}}]{Berkefeld_2012}
{Berkefeld }, T., {Schmidt}, D., {Soltau}, D., {von der L{\"u}he}, O., \&
  {Heidecke}, F. 2012, Astronomische Nachrichten, 333, 863

\bibitem[{{Bloomfield} {et~al.}(2007){Bloomfield}, {Solanki}, {Lagg},
  {Borrero}, \& {Cally}}]{Bloomfield_2007}
{Bloomfield}, D.~S., {Solanki}, S.~K., {Lagg}, A., {Borrero}, J.~M., \&
  {Cally}, P.~S. 2007, \aap, 469, 1155

\bibitem[{{Bonet} {et~al.}(2010){Bonet}, {M{\'a}rquez}, {S{\'a}nchez Almeida},
  {Palacios}, {Mart{\'{\i}}nez Pillet}, {Solanki}, {del Toro Iniesta},
  {Domingo}, {Berkefeld}, {Schmidt}, {Gandorfer}, {Barthol}, \&
  {Kn{\"o}lker}}]{bonet2010imax}
{Bonet}, J.~A., {M{\'a}rquez}, I., {S{\'a}nchez Almeida}, J., {et~al.} 2010,
  \apjl, 723, L139

\bibitem[{{Borrero} {et~al.}(2016){Borrero}, {Asensio Ramos}, {Collados},
  {Schlichenmaier}, {Balthasar}, {Franz}, {Rezaei}, {Kiess}, {Orozco
  Su{\'a}rez}, {Pastor}, {Berkefeld}, {von der L{\"u}he}, {Schmidt}, {Schmidt},
  {Sigwarth}, {Soltau}, {Volkmer}, {Waldmann}, {Denker}, {Hofmann}, {Staude},
  {Strassmeier}, {Feller}, {Lagg}, {Solanki}, {Sobotka}, \&
  {Nicklas}}]{Borrero_2016}
{Borrero}, J.~M., {Asensio Ramos}, A., {Collados}, M., {et~al.} 2016, \aap,
  596, A2

\bibitem[{{Borrero} {et~al.}(2003){Borrero}, {Bellot Rubio}, {Barklem}, \& {del
  Toro Iniesta}}]{Borrero_2003}
{Borrero}, J.~M., {Bellot Rubio}, L.~R., {Barklem}, P.~S., \& {del Toro
  Iniesta}, J.~C. 2003, \aap, 404, 749

\bibitem[{{Borrero} {et~al.}(2007){Borrero}, {Bellot Rubio}, \&
  {M{\"u}ller}}]{borrero2007ncp}
{Borrero}, J.~M., {Bellot Rubio}, L.~R., \& {M{\"u}ller}, D.~A.~N. 2007, \apjl,
  666, L133

\bibitem[{{Borrero} {et~al.}(2017){Borrero}, {Franz}, {Schlichenmaier},
  {Collados}, \& {Asensio Ramos}}]{Borrero_2017}
{Borrero}, J.~M., {Franz}, M., {Schlichenmaier}, R., {Collados}, M., \&
  {Asensio Ramos}, A. 2017, \aap, 601, L8

\bibitem[{{Borrero} {et~al.}(2005){Borrero}, {Lagg}, {Solanki}, \&
  {Collados}}]{Borrero_2005}
{Borrero}, J.~M., {Lagg}, A., {Solanki}, S.~K., \& {Collados}, M. 2005, \aap,
  436, 333

\bibitem[{{Borrero} {et~al.}(2010){Borrero}, {Mart{\'{\i}}nez-Pillet},
  {Schlichenmaier}, {Solanki}, {Bonet}, {del Toro Iniesta}, {Schmidt},
  {Barthol}, {Gandorfer}, {Domingo}, \& {Kn{\"o}lker}}]{Borrero_2010}
{Borrero}, J.~M., {Mart{\'{\i}}nez-Pillet}, V., {Schlichenmaier}, R., {et~al.}
  2010, \apjl, 723, L144

\bibitem[{{Borrero} {et~al.}(2013){Borrero}, {Mart{\'{\i}}nez Pillet},
  {Schmidt}, {Quintero Noda}, {Bonet}, {del Toro Iniesta}, \& {Bellot
  Rubio}}]{Borrero_2013}
{Borrero}, J.~M., {Mart{\'{\i}}nez Pillet}, V., {Schmidt}, W., {et~al.} 2013,
  \apj, 768, 69

\bibitem[{{Borrero} \& {Solanki}(2010)}]{borrero2010ncp}
{Borrero}, J.~M. \& {Solanki}, S.~K. 2010, \apj, 709, 349

\bibitem[{{Borrero} {et~al.}(2004){Borrero}, {Solanki}, {Bellot Rubio}, {Lagg},
  \& {Mathew}}]{Borrero_2004}
{Borrero}, J.~M., {Solanki}, S.~K., {Bellot Rubio}, L.~R., {Lagg}, A., \&
  {Mathew}, S.~K. 2004, \aap, 422, 1093

\bibitem[{{Borrero} {et~al.}(2006){Borrero}, {Solanki}, {Lagg},
  {Socas-Navarro}, \& {Lites}}]{borrero2006}
{Borrero}, J.~M., {Solanki}, S.~K., {Lagg}, A., {Socas-Navarro}, H., \&
  {Lites}, B. 2006, \aap, 450, 383

\bibitem[{{Cabrera Solana} {et~al.}(2008){Cabrera Solana}, {Bellot Rubio},
  {Borrero}, \& {Del Toro Iniesta}}]{cabrera2008}
{Cabrera Solana}, D., {Bellot Rubio}, L.~R., {Borrero}, J.~M., \& {Del Toro
  Iniesta}, J.~C. 2008, \aap, 477, 273

\bibitem[{{Chandrasekhar} \& {Breen}(1946)}]{chandra1946hminus}
{Chandrasekhar}, S. \& {Breen}, F.~H. 1946, \apj, 104, 430

\bibitem[{{Collados} {et~al.}(2012){Collados}, {L{\'o}pez}, {P{\'a}ez},
  {Hern{\'a}ndez}, {Reyes}, {Calcines}, {Ballesteros}, {D{\'{\i}}az}, {Denker},
  {Lagg}, {Schlichenmaier}, {Schmidt}, {Solanki}, {Strassmeier}, {von der
  L{\"u}he}, \& {Volkmer}}]{Collados_2012}
{Collados}, M., {L{\'o}pez}, R., {P{\'a}ez}, E., {et~al.} 2012, Astronomische
  Nachrichten, 333, 872

\bibitem[{{Danilovic} {et~al.}(2015){Danilovic}, {Cameron}, \&
  {Solanki}}]{Danilovic_2015}
{Danilovic}, S., {Cameron}, R.~H., \& {Solanki}, S.~K. 2015, \aap, 574, A28

\bibitem[{{del Toro Iniesta} {et~al.}(2001){del Toro Iniesta}, {Bellot Rubio},
  \& {Collados}}]{Toro_Iniesta_2001}
{del Toro Iniesta}, J.~C., {Bellot Rubio}, L.~R., \& {Collados}, M. 2001,
  \apjl, 549, L139

\bibitem[{{del Toro Iniesta} \& {Ruiz Cobo}(2016)}]{Iniesta_Cobo_2016}
{del Toro Iniesta}, J.~C. \& {Ruiz Cobo}, B. 2016, Living Reviews in Solar
  Physics, 13, 4

\bibitem[{{Franz} {et~al.}(2016){Franz}, {Collados}, {Bethge},
  {Schlichenmaier}, {Borrero}, {Schmidt}, {Lagg}, {Solanki}, {Berkefeld},
  {Kiess}, {Rezaei}, {Schmidt}, {Sigwarth}, {Soltau}, {Volkmer}, {von der
  Luhe}, {Waldmann}, {Orozco}, {Pastor Yabar}, {Denker}, {Balthasar}, {Staude},
  {Hofmann}, {Strassmeier}, {Feller}, {Nicklas}, {Kneer}, \&
  {Sobotka}}]{Franz_2016}
{Franz}, M., {Collados}, M., {Bethge}, C., {et~al.} 2016, \aap, 596, A4

\bibitem[{{Franz} \& {Schlichenmaier}(2013)}]{Franz_2013}
{Franz}, M. \& {Schlichenmaier}, R. 2013, \aap, 550, A97

\bibitem[{{Gingerich} {et~al.}(1971){Gingerich}, {Noyes}, {Kalkofen}, \&
  {Cuny}}]{hsra1974}
{Gingerich}, O., {Noyes}, R.~W., {Kalkofen}, W., \& {Cuny}, Y. 1971, \solphys,
  18, 347

\bibitem[{{Holweger} \& {Mueller}(1974)}]{holweger1974}
{Holweger}, H. \& {Mueller}, E.~A. 1974, \solphys, 39, 19

\bibitem[{{Ichimoto} {et~al.}(2007){Ichimoto}, {Shine}, {Lites}, {Kubo},
  {Shimizu}, {Suematsu}, {Tsuneta}, {Katsukawa}, {Tarbell}, {Title}, {Nagata},
  {Yokoyama}, \& {Shimojo}}]{Ichimoto_2007}
{Ichimoto}, K., {Shine}, R.~A., {Lites}, B., {et~al.} 2007, \pasj, 59, S593

\bibitem[{{Ichimoto} {et~al.}(2008){Ichimoto}, {Tsuneta}, {Suematsu},
  {Katsukawa}, {Shimizu}, {Lites}, {Kubo}, {Tarbell}, {Shine}, {Title}, \&
  {Nagata}}]{ichimoto2008ncp}
{Ichimoto}, K., {Tsuneta}, S., {Suematsu}, Y., {et~al.} 2008, \aap, 481, L9

\bibitem[{{Jafarzadeh} {et~al.}(2015){Jafarzadeh}, {Rouppe van der Voort}, \&
  {de la Cruz Rodr{\'{\i}}guez}}]{Jafarzadeh_2015}
{Jafarzadeh}, S., {Rouppe van der Voort}, L., \& {de la Cruz Rodr{\'{\i}}guez},
  J. 2015, \apj, 810, 54

\bibitem[{{Khomenko} {et~al.}(2003){Khomenko}, {Collados}, {Solanki}, {Lagg},
  \& {Trujillo Bueno}}]{Khomenko_2003}
{Khomenko}, E.~V., {Collados}, M., {Solanki}, S.~K., {Lagg}, A., \& {Trujillo
  Bueno}, J. 2003, \aap, 408, 1115

\bibitem[{{Lagg} {et~al.}(2016){Lagg}, {Solanki}, {Doerr}, {Mart{\'{\i}}nez
  Gonz{\'a}lez}, {Riethm{\"u}ller}, {Collados Vera}, {Schlichenmaier}, {Orozco
  Su{\'a}rez}, {Franz}, {Feller}, {Kuckein}, {Schmidt}, {Asensio Ramos},
  {Pastor Yabar}, {von der L{\"u}he}, {Denker}, {Balthasar}, {Volkmer},
  {Staude}, {Hofmann}, {Strassmeier}, {Kneer}, {Waldmann}, {Borrero},
  {Sobotka}, {Verma}, {Louis}, {Rezaei}, {Soltau}, {Berkefeld}, {Sigwarth},
  {Schmidt}, {Kiess}, \& {Nicklas}}]{Lagg_2016}
{Lagg}, A., {Solanki}, S.~K., {Doerr}, H.-P., {et~al.} 2016, \aap, 596, A6

\bibitem[{{Lagg} {et~al.}(2010){Lagg}, {Solanki}, {Riethm{\"u}ller},
  {Mart{\'{\i}}nez Pillet}, {Sch{\"u}ssler}, {Hirzberger}, {Feller}, {Borrero},
  {Schmidt}, {del Toro Iniesta}, {Bonet}, {Barthol}, {Berkefeld}, {Domingo},
  {Gandorfer}, {Kn{\"o}lker}, \& {Title}}]{lagg2010imax}
{Lagg}, A., {Solanki}, S.~K., {Riethm{\"u}ller}, T.~L., {et~al.} 2010, \apjl,
  723, L164

\bibitem[{{Landi Degl'Innocenti} \& {Landi Degl'Innocenti}(1981)}]{landi81ncp}
{Landi Degl'Innocenti}, E. \& {Landi Degl'Innocenti}, M. 1981, Nuovo Cimento B
  Serie, 62, 1

\bibitem[{{Landi Degl'Innocenti} \& {Landolfi}(1983)}]{landi83ncp}
{Landi Degl'Innocenti}, E. \& {Landolfi}, M. 1983, \solphys, 87, 221

\bibitem[{{Landolfi} \& {Landi Degl'Innocenti}(1996)}]{landolfi1996ncp}
{Landolfi}, M. \& {Landi Degl'Innocenti}, E. 1996, \solphys, 164, 191

\bibitem[{{Livingston} \& {Wallace}(1991)}]{Livingston_Wallace_1991}
{Livingston}, W. \& {Wallace}, L. 1991, {An atlas of the solar spectrum in the
  infrared from 1850 to 9000 cm-1 (1.1 to 5.4 micrometer)}

\bibitem[{{Louis} {et~al.}(2009){Louis}, {Bellot Rubio}, {Mathew}, \&
  {Venkatakrishnan}}]{louis2009}
{Louis}, R.~E., {Bellot Rubio}, L.~R., {Mathew}, S.~K., \& {Venkatakrishnan},
  P. 2009, \apjl, 704, L29

\bibitem[{{Mart{\'{\i}}nez Gonz{\'a}lez} {et~al.}(2008){Mart{\'{\i}}nez
  Gonz{\'a}lez}, {Asensio Ramos}, {L{\'o}pez Ariste}, \& {Manso
  Sainz}}]{marian2008qs}
{Mart{\'{\i}}nez Gonz{\'a}lez}, M.~J., {Asensio Ramos}, A., {L{\'o}pez Ariste},
  A., \& {Manso Sainz}, R. 2008, \aap, 479, 229

\bibitem[{{Mart{\'{\i}}nez Gonz{\'a}lez} {et~al.}(2012){Mart{\'{\i}}nez
  Gonz{\'a}lez}, {Bellot Rubio}, {Solanki}, {Mart{\'{\i}}nez Pillet}, {Del Toro
  Iniesta}, {Barthol}, \& {Schmidt}}]{marian2012sirjump}
{Mart{\'{\i}}nez Gonz{\'a}lez}, M.~J., {Bellot Rubio}, L.~R., {Solanki}, S.~K.,
  {et~al.} 2012, \apjl, 758, L40

\bibitem[{{Mart{\'{\i}}nez Gonz{\'a}lez} {et~al.}(2016){Mart{\'{\i}}nez
  Gonz{\'a}lez}, {Pastor Yabar}, {Lagg}, {Asensio Ramos}, {Collados},
  {Solanki}, {Balthasar}, {Berkefeld}, {Denker}, {Doerr}, {Feller}, {Franz},
  {Gonz{\'a}lez Manrique}, {Hofmann}, {Kneer}, {Kuckein}, {Louis}, {von der
  L{\"u}he}, {Nicklas}, {Orozco}, {Rezaei}, {Schlichenmaier}, {Schmidt},
  {Schmidt}, {Sigwarth}, {Sobotka}, {Soltau}, {Staude}, {Strassmeier}, {Verma},
  {Waldman}, \& {Volkmer}}]{marian2016gregor}
{Mart{\'{\i}}nez Gonz{\'a}lez}, M.~J., {Pastor Yabar}, A., {Lagg}, A., {et~al.}
  2016, \aap, 596, A5

\bibitem[{{Mart{\'{\i}}nez Pillet} {et~al.}(2011){Mart{\'{\i}}nez Pillet}, {Del
  Toro Iniesta}, \& {Quintero Noda}}]{Martinez_Pillet_2011}
{Mart{\'{\i}}nez Pillet}, V., {Del Toro Iniesta}, J.~C., \& {Quintero Noda}, C.
  2011, \aap, 530, A111

\bibitem[{{M{\"u}ller} {et~al.}(2002){M{\"u}ller}, {Schlichenmaier}, {Steiner},
  \& {Stix}}]{mueller2002ncp}
{M{\"u}ller}, D.~A.~N., {Schlichenmaier}, R., {Steiner}, O., \& {Stix}, M.
  2002, \aap, 393, 305

\bibitem[{{Nave} {et~al.}(1994){Nave}, {Johansson}, {Learner}, {Thorne}, \&
  {Brault}}]{Nave_1994}
{Nave}, G., {Johansson}, S., {Learner}, R.~C.~M., {Thorne}, A.~P., \& {Brault},
  J.~W. 1994, \apjs, 94, 221

\bibitem[{{Pozuelo} {et~al.}(2016){Pozuelo}, {Bellot Rubio}, \& {de la Cruz
  Rodr{\'{\i}}guez}}]{Pozuelo_2016}
{Pozuelo}, S.~E., {Bellot Rubio}, L.~R., \& {de la Cruz Rodr{\'{\i}}guez}, J.
  2016, \apj, 832, 170

\bibitem[{{Press} {et~al.}(1986){Press}, {Flannery}, \&
  {Teukolsky}}]{Press_1986}
{Press}, W.~H., {Flannery}, B.~P., \& {Teukolsky}, S.~A. 1986, {Numerical
  recipes. The art of scientific computing}

\bibitem[{{Quintero Noda} {et~al.}(2014{\natexlab{a}}){Quintero Noda},
  {Borrero}, {Orozco Su{\'a}rez}, \& {Ruiz Cobo}}]{Quintero_Noda_2014a}
{Quintero Noda}, C., {Borrero}, J.~M., {Orozco Su{\'a}rez}, D., \& {Ruiz Cobo},
  B. 2014{\natexlab{a}}, \aap, 569, A73

\bibitem[{{Quintero Noda} {et~al.}(2013){Quintero Noda}, {Mart{\'{\i}}nez
  Pillet}, {Borrero}, \& {Solanki}}]{Quintero_Noda_2013}
{Quintero Noda}, C., {Mart{\'{\i}}nez Pillet}, V., {Borrero}, J.~M., \&
  {Solanki}, S.~K. 2013, \aap, 558, A30

\bibitem[{{Quintero Noda} {et~al.}(2014{\natexlab{b}}){Quintero Noda}, {Ruiz
  Cobo}, \& {Orozco Su{\'a}rez}}]{Quintero_Noda_2014b}
{Quintero Noda}, C., {Ruiz Cobo}, B., \& {Orozco Su{\'a}rez}, D.
  2014{\natexlab{b}}, \aap, 566, A139

\bibitem[{{Rimmele} {et~al.}(2015){Rimmele}, {McMullin}, {Warner}, {Craig},
  {Woeger}, {Tritschler}, {Cassini}, {Kuhn}, {Lin}, {Schmidt}, {Berukoff},
  {Reardon}, {Goode}, {Knoelker}, {Rosner}, {Mathioudakis}, \& {DKIST
  TEAM}}]{Rimmele_2015}
{Rimmele}, T., {McMullin}, J., {Warner}, M., {et~al.} 2015, IAU General
  Assembly, 22, 2255176

\bibitem[{{Rueedi} {et~al.}(1998){Rueedi}, {Solanki}, {Keller}, \&
  {Frutiger}}]{ruedi1998}
{Rueedi}, I., {Solanki}, S.~K., {Keller}, C.~U., \& {Frutiger}, C. 1998, \aap,
  338, 1089

\bibitem[{{Ruiz Cobo} \& {del Toro Iniesta}(1992)}]{Cobo_Iniesta_1992}
{Ruiz Cobo}, B. \& {del Toro Iniesta}, J.~C. 1992, \apj, 398, 375

\bibitem[{{Sainz Dalda} {et~al.}(2012){Sainz Dalda}, {Mart{\'{\i}}nez-Sykora},
  {Bellot Rubio}, \& {Title}}]{Sainz_Dalda_2012}
{Sainz Dalda}, A., {Mart{\'{\i}}nez-Sykora}, J., {Bellot Rubio}, L., \&
  {Title}, A. 2012, \apj, 748, 38

\bibitem[{{Sanchez Almeida} {et~al.}(1988){Sanchez Almeida}, {Collados}, \&
  {del Toro Iniesta}}]{jorge1988ncp}
{Sanchez Almeida}, J., {Collados}, M., \& {del Toro Iniesta}, J.~C. 1988, \aap,
  201, L37

\bibitem[{{Sanchez Almeida} {et~al.}(1989){Sanchez Almeida}, {Collados}, \&
  {del Toro Iniesta}}]{jorge1989ncp}
{Sanchez Almeida}, J., {Collados}, M., \& {del Toro Iniesta}, J.~C. 1989, \aap,
  222, 311

\bibitem[{{Sanchez Almeida} \& {Lites}(1992)}]{Sanchez_Almeida_1992}
{Sanchez Almeida}, J. \& {Lites}, B.~W. 1992, \apj, 398, 359

\bibitem[{{Schlichenmaier} \& {Collados}(2002)}]{rolf2002ncp1}
{Schlichenmaier}, R. \& {Collados}, M. 2002, \aap, 381, 668

\bibitem[{{Schlichenmaier} {et~al.}(2002){Schlichenmaier}, {M{\"u}ller},
  {Steiner}, \& {Stix}}]{rolf2002ncp2}
{Schlichenmaier}, R., {M{\"u}ller}, D.~A.~N., {Steiner}, O., \& {Stix}, M.
  2002, \aap, 381, L77

\bibitem[{{Schmidt} {et~al.}(2012){Schmidt}, {von der L{\"u}he}, {Volkmer},
  {Denker}, {Solanki}, {Balthasar}, {Bello Gonz{\'a}lez}, {Berkefeld},
  {Collados Vera}, {Hofmann}, {Kneer}, {Lagg}, {Puschmann}, {Schmidt},
  {Sobotka}, {Soltau}, \& {Strassmeier}}]{Schmidt_2012}
{Schmidt}, W., {von der L{\"u}he}, O., {Volkmer}, R., {et~al.} 2012, in
  Astronomical Society of the Pacific Conference Series, Vol. 463, Second
  ATST-EAST Meeting: Magnetic Fields from the Photosphere to the Corona., ed.
  T.~R. {Rimmele}, A.~{Tritschler}, F.~{W{\"o}ger}, M.~{Collados Vera},
  H.~{Socas-Navarro}, R.~{Schlichenmaier}, M.~{Carlsson}, T.~{Berger},
  A.~{Cadavid}, P.~R. {Gilbert}, P.~R. {Goode}, \& M.~{Kn{\"o}lker}, 365

\bibitem[{{Shaltout} \& {Ichimoto}(2015)}]{shalout2015sirjump}
{Shaltout}, A.~M.~K. \& {Ichimoto}, K. 2015, \pasj, 67, 27

\bibitem[{{Socas-Navarro} \& {Manso Sainz}(2005)}]{Socas_Navarro_2005}
{Socas-Navarro}, H. \& {Manso Sainz}, R. 2005, \apjl, 620, L71

\bibitem[{{Solanki} \& {Montavon}(1993)}]{solanki1993ncp}
{Solanki}, S.~K. \& {Montavon}, C.~A.~P. 1993, \aap, 275, 283

\bibitem[{{Stenflo} {et~al.}(1984){Stenflo}, {Solanki}, {Harvey}, \&
  {Brault}}]{stenflo1984da}
{Stenflo}, J.~O., {Solanki}, S., {Harvey}, J.~W., \& {Brault}, J.~W. 1984,
  \aap, 131, 333

\bibitem[{{Stix}(2004)}]{Stix_2004}
{Stix}, M. 2004, {The sun : an introduction}

\bibitem[{{Vernazza} {et~al.}(1981){Vernazza}, {Avrett}, \&
  {Loeser}}]{vernazza1981}
{Vernazza}, J.~E., {Avrett}, E.~H., \& {Loeser}, R. 1981, \apjs, 45, 635

\bibitem[{{Viticchi{\'e}}(2012)}]{Viticchie_2012}
{Viticchi{\'e}}, B. 2012, \apjl, 747, L36

\bibitem[{{Wedemeyer-B{\"o}hm} {et~al.}(2012){Wedemeyer-B{\"o}hm}, {Scullion},
  {Steiner}, {Rouppe van der Voort}, {de La Cruz Rodriguez}, {Fedun}, \&
  {Erd{\'e}lyi}}]{wedemeyer2012vortex}
{Wedemeyer-B{\"o}hm}, S., {Scullion}, E., {Steiner}, O., {et~al.} 2012, \nat,
  486, 505

\end{thebibliography}

\clearpage
\newpage

\appendix
\section{}

Inversion results of selected pixels with the red contour in Fig.~\ref{fig:continuummap_contour}. The black curves show the measured Stokes profiles, while the red/blue curves show
the inversion results for the one/two component inversion, respectively.

\begin{figure}[h]
\resizebox{\hsize}{!}{\includegraphics{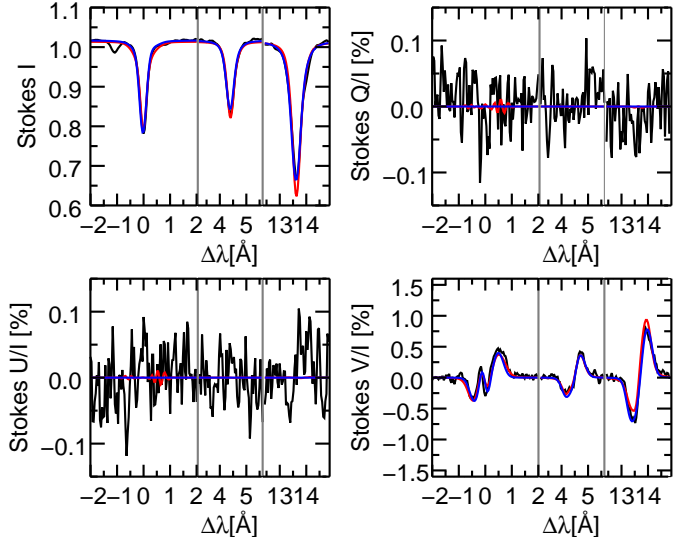}}
\caption{First example of observed (black), fitted with the one-component inversion (red; Sects.~\ref{section:one_component},~\ref{section:one_component_result}),
and with the two-component inversion (blue; Sects.~\ref{section:two_component},~\ref{section:two_component_result}), Stokes profiles in a pixel of the red contour in 
Fig.~\ref{fig:continuummap_contour}.\label{fig:appendix_first}}
\end{figure}

\begin{figure}[h]
\resizebox{\hsize}{!}{\includegraphics{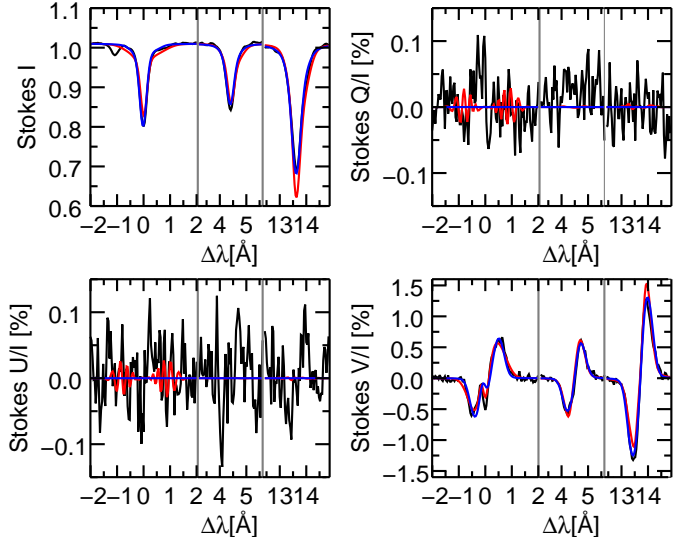}}
\caption{As in Fig.~\ref{fig:appendix_first} but for a second pixel of the red region in Fig.~\ref{fig:continuummap_contour}.\label{fig:appendix_second}}
\end{figure}

\begin{figure}[h]
\resizebox{\hsize}{!}{\includegraphics{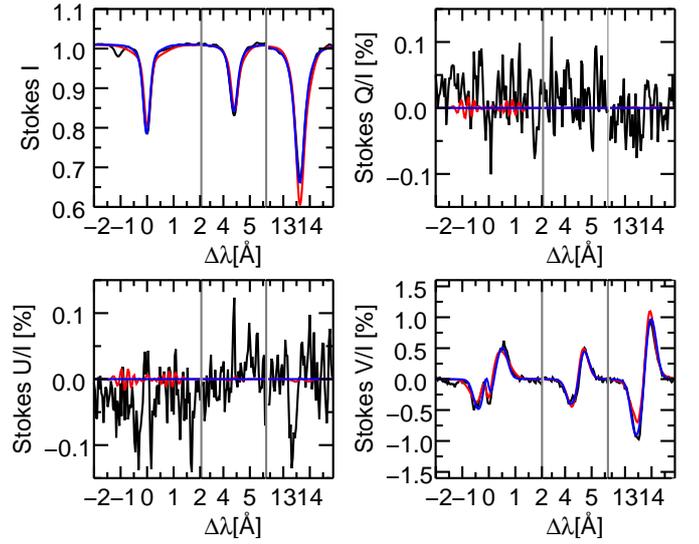}}
\caption{As in Fig.~\ref{fig:appendix_first} but for a third pixel of the red region in Fig.~\ref{fig:continuummap_contour}.\label{fig:appendix_third}}
\end{figure}

\begin{figure}[h]
\resizebox{\hsize}{!}{\includegraphics{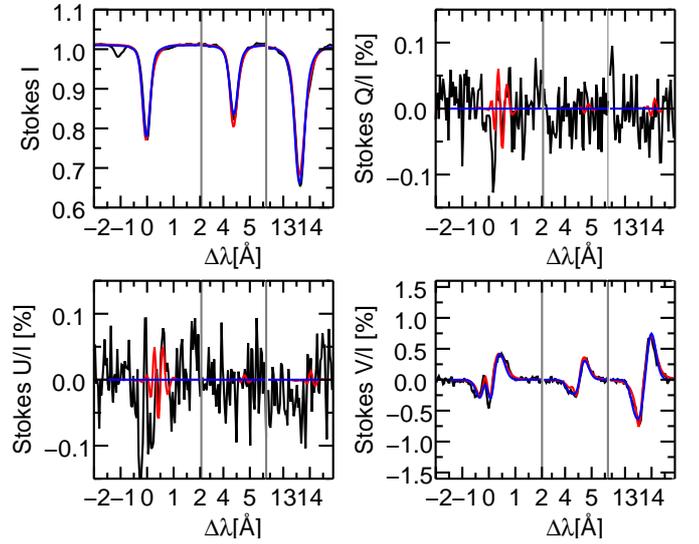}}
\caption{As in Fig.~\ref{fig:appendix_first} but for a fourth pixel of the red region in Fig.~\ref{fig:continuummap_contour}.\label{fig:appendix_fourth}}
\end{figure}

\end{document}